\documentclass[12pt,a4paper]{article}% Vancouver Reference 

\usepackage{amsmath}
\usepackage{mathtools}
\usepackage{enumitem}
\usepackage{caption}
\usepackage{hyperref}
\usepackage{amsfonts}
\usepackage{abstract}
\usepackage{authblk} % author and affiliation for arXiv publication

\DeclareMathAlphabet\mathbfcal{OMS}{cmsy}{b}{n}

\begin{document}

\title{Neural Networks for Constitutive Modeling - From Universal Function Approximators to Advanced Models and the Integration of Physics}

\author[1]{Johannes Dornheim\thanks{corresponding author, johannes.dornheim@mailbox.org}}
\author[2]{Lukas Morand\thanks{corresponding author, lukas.morand@iwm.fraunhofer.de}}
\author[2]{Hemanth Janarthanam Nallani\thanks{hemanth.janarthanam@iwm.fraunhofer.de}}
\author[2]{Dirk Helm\thanks{dirk.helm@iwm.fraunhofer.de}}

\affil[1]{Institute for Applied Materials, Karlsruhe Institute of Technology, Germany}

\affil[2]{Fraunhofer Institute for Mechanics of Materials IWM, Freiburg, Germany}

\maketitle

\begin{abstract}
Analyzing and modeling the constitutive behavior of materials is a core area in materials sciences and a prerequisite for conducting numerical simulations in which the material behavior plays a central role. Constitutive models have been developed since the beginning of the 19th century and are still under constant development. Besides physics-motivated and phenomenological models, during the last decades, the field of constitutive modeling was enriched by the development of machine learning-based constitutive models, especially by using neural networks. The latter is the focus of the present review paper, which aims to give an overview of neural networks-based constitutive models from a methodical perspective. The review summarizes and compares numerous conceptually different neural networks-based approaches for constitutive modeling including neural networks used as universal function approximators, advanced neural network models and neural network approaches with integrated physical knowledge. The upcoming of these methods is in-turn closely related to advances in the area of computer sciences, what further adds a chronological aspect to this review. We conclude the review paper with important challenges in the field of learning constitutive relations that need to be tackled in the near future.
\end{abstract}

\section{Introduction and Background}\label{intro}

\subsection{Motivation}

In continuum mechanics, we distinguish between universal relations (e.g. balance laws) and constitutive relations that describe the behavior of a specific class of materials. Using the theory of materials (see for example \cite{haupt2013continuum}), useful principles can be formulated that form the basis for development of these constitutive relations. Within this framework, constitutive models are developed based on phenomenological observations and/or physical knowledge, often related to physically motivated state variables like dislocation density in metallic materials. In addition, constitutive relations can be formulated in accordance to the second law of thermodynamics which enforces thermodynamically consistent models. Independent of the modeling approach, material models contain material dependent parameters, which are to be calibrated using experiments, potentially accompanied by  numerical simulations like virtual testing techniques \cite{zhang2016virtual}. Typically, the advantage of physically motivated models are the reduced number of material dependent parameters that need to be determined from experiments. 

Over the past few decades, a new approach based on supervised machine learning techniques for modeling non-linear material behavior has emerged, specifically using neural networks. In general, compared to complex physics-based models, machine learning-based models can be executed in near real time, which enables extensive accelerated numerical simulations. This in-turn can lead to a breakthrough in engineering, as it allows for more accurate and detailed numerical simulations on component and process level. Neural networks-based approaches offer a significant advantage over other supervised learning models by virtue of their ability to represent any continuous functional relation \cite{cybenko1989approximation, leshno1993multilayer} and thus model arbitrary complex material behavior. However, the large number of parameters that need to be calibrated during the neural network training is a shortcoming.

Besides constitutive modeling, the potential of using machine learning models is actively being investigated across multiple fields of computational material science. Among others, these methods are used in computational design of materials \cite{liu2015predictive,bessa2017framework,tran2020active,iraki2021multi,morand2022efficient}, in design of processes \cite{dornheim2020model,dornheim2018multiobjective, liu2020reinforcement}, in development of digital twins \cite{chinesta2020virtual} and soft sensors \cite{kadlec2009data}, and in multi-scale simulations and homogenization schemes. Numerical prediction in these applications often involve frequent execution of simulations with variations of model parameters \cite{wang2018multiscale}. The central limiting factors of classical modeling approach are the complexity of constitutive models and computational performance of the underlying simulations. In order to improve the computational performance, surrogate models (i.e. learned simulations or constitutive models) can be set up on various levels.

Surrogate neural networks are often proposed at structural/component level \cite{zhang2019deep, koeppe2019efficient, im2021surrogate, stoffel2020artificial}, at continuum level across multiple length scales \cite{wang2018multiscale} to replace conventional constitutive models, or at atomic and molecular level \cite{noe2020machine}. As opposed to structural surrogates, learned constitutive models, once trained, are in principle applicable across different structures. Methods to train such constitutive neural network models (either directly or indirectly) are proposed by various authors. We outline and classify these training approaches in Section \ref{methods_classification}.

The idea of using neural networks to learn constitutive relations dates back to the pioneering work of Ghaboussi et al. \cite{ghaboussi1991knowledge} in the early 90s. A first short but general review summarizing the use of neural networks in computational mechanics, including constitutive modeling, was published in 1996 by Yagawa and Okuda \cite{yagawa1996neural}. Due to the recent heightened interest in this research field, a variety of review and survey papers have been published. Notably reviews focusing on the use of machine learning methods for materials discovery and design \cite{liu2017materials, guo2021artificial}, for  engineering of materials, processes, and structures \cite{dimiduk2018perspectives}, for multi-scale modeling \cite{peng2021multiscale} and for meta materials design \cite{jiao2021artificial} have been presented. A broad review on the application of machine learning methods in continuum mechanics has been put forth by Bock et al. \cite{bock2019review}. Pertaining to constitutive models, reviews focused on speicific material classes like composites \cite{liu2021review}, soils \cite{zhang2021state}, alloys \cite{hart2021machine}, and sheet metals specific to forming processes \cite{lourencco2022use}.

This contribution, on the contrary, is a general characterization and classification of neural network methods for learning constitutive behavior. This review also has a chronological aspect, as the development of these methods evolved with advancements in the field of machine learning, such as the development of convolutional neural networks (CNNs) and recurrent neural networks (RNNs), see \cite{goodfellow2016deep} for both, as well as physics-informed neural networks (PINNs), see \cite{raissi2017physics} (although such approaches have already been proposed in the late 90s, for example in \cite{lagaris1998artificial}). The purpose of PINNs in \cite{raissi2017physics} however, is basically to solve differential equations on a domain for given boundary values, which is primarily not directly transferable for learning constitutive relations. However, the combination of the physics-informed part of PINNs (which can be understood as an encoding of a physical law described by a differential equation) and the neural network part (which predicts the quantities of interest being fully differentiable with respect to all input variables by using automatic differentiation \cite{baydin2018automatic}), can be adjusted and leveraged for learning constitutive relations.

\begin{figure}[h]
  	\centering
	\includegraphics[width=1\linewidth]{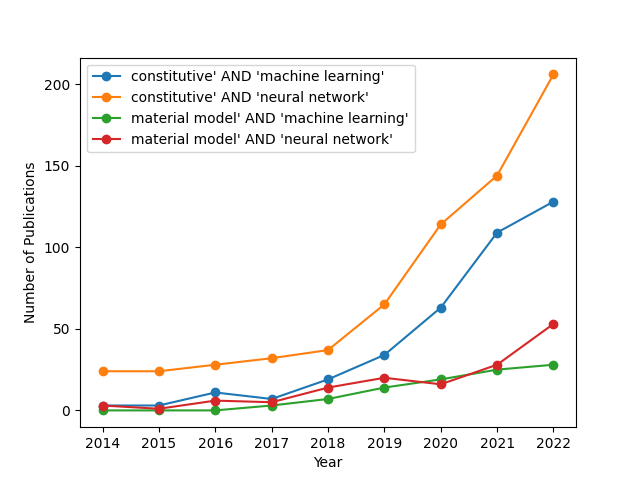}
	\caption{Number of annual publications (articles, preprints, chapters or proceedings) that contain the 
	  combination of the following keywords within the title or the abstract: constitutive and machine 
	  learning, constitutive and neural network, material model and neural network, material model and 
	  machine learning. The data is gathered from \url{app.dimensions.ai}.}
	\label{fig:paper-count}
\end{figure}

The field of research on the application of  machine learning to constitutive modeling is very active and constantly expanding. This is reflected in the increase of annually published papers as shown in Figure \ref{fig:paper-count}. As novel papers are published almost on a weekly basis, we limit this review to works that have been published before 2022 with a few important contributions from 2022. From the methodic viewpoint, we focus solely on approaches that utilize neural networks for learning constitutive relations. Methods beyond our scope are structure-level surrogates \cite{samaniego2020energy, nguyen2020deep, fuhg2021mixed, fernandez2020application}, data-driven solvers \cite{kirchdoerfer2016data, ibanez2017data, ibanez2018manifold, eggersmann2021model, karapiperis2021data}, constitutive model free approaches \cite{asteris2017feed, nguyen2018data, wang2021non}, and constitutive modeling using symbolic machine learning \cite{versino2017data}, random forests \cite{reimann2019modeling} and spline interpolation \cite{crespo2017wypiwyg, romero2017determination, latorre2020experimental} methods.

\subsection{Classification of Neural Networks-Based Constitutive Modeling Approaches}\label{methods_classification}
\begin{figure}[h]
	\centering
	\includegraphics[width=0.7\linewidth]{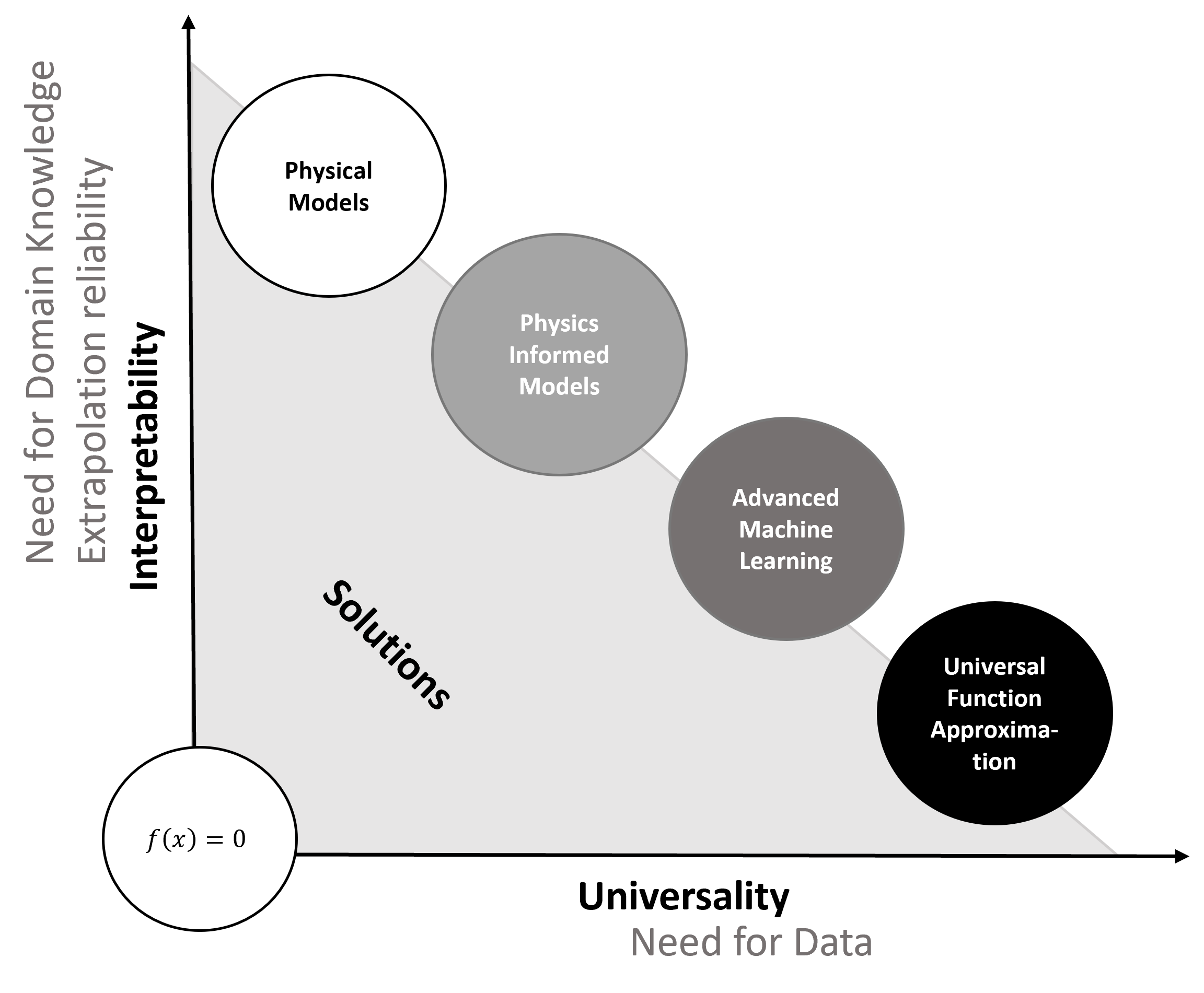}
	\caption{Methodical spectrum of named constitutive modeling approaches and an outline of the central trade off of these models}
	\label{fig:methods_overview}
\end{figure}

The numerous neural networks-based constitutive modeling approaches proposed in literature can be distributed along an axis, as depicted in Figure \ref{fig:methods_overview}, from purely physics-based to universal function approximation models. The classical feedforward neural networks (FFNNs) is placed at one extremum of the spectrum. FFNNs can be used to learn constitutive models purely from data without any knowledge of the underlying physical relations. Using the universal function approximation theorem  \cite{cybenko1989approximation,leshno1993multilayer}, it has been shown that, given a sufficient amount of model parameters, FFNNs can in principle be parameterized to represent any continuous functional relation. In this regard, neural networks can be used to learn functional relations that typically originate from material model formulations. However, the need for large set of labeled data and the lack of interpretability are few of the drawbacks of FFNNs in its basic form. Other types of neural networks fill the gap between the purely data-driven neural network models and the conventional physical constitutive modeling approaches. As depicted in Figure \ref{fig:methods_overview}, we associate five general characteristics to the methods with in the spectrum.
\begin{enumerate}
	\item % expert knowledge <-> sufficient data
The basic requirement for building a physical constitutive model is sufficient \textbf{domain knowledge}, more specifically, physical knowledge about the material behavior.
\item
In contrast, for a data-driven modeling approach there is a \textbf{need for data}, covering the full spectrum of the material behavior. 
\item % bias <-> unbiased
While, due to imperfections and simplifying assumptions, physical models are usually biased, trained FFNNs, under the assumption of sufficient data and enough model capacity, are unbiased. A \textbf{universal} estimation of the true material behavior is possible.
\item % application range limited by data, dangerous extrapolation <-> safe-extrapolation but - often limited application range
Purely data-driven models learn by minimizing the difference between predictions and ground-truth data and usually have no incentive to \textbf{extrapolate} beyond the ground truth. Combined with the non-linearity of such models, this typically leads to a highly non-linear model behavior outside of the training region. The incorporation of physical knowledge can be used to constrain the model and thereby improves model robustness, especially when extrapolating.
\item % black-box <-> physically explainable
Another important aspect is the \textbf{interpretability} of the model. Machine learning models for universal approximation of any non-trivial function involve a large set of parameters with no physical meaning and complicated influence on the approximations. Such models have to be seen as black boxes. Physical models, on the other hand, are characterized by the fact that the parameters and their impact are well-defined. 
\end{enumerate}
There are advanced techniques available in neural networks-based constitutive modeling that surpass the use of simple FFNNs. In the following, we distinguish between two classes of approaches that incorporate domain knowledge into the neural networks to improve data efficiency, explainability, and extrapolation capabilities of the learned constitutive model. The first class of approach is based on the use of certain pre-structured layer and advanced neural network architectures to integrate knowledge about the temporal or geometric structure of the problem and data at hand. This includes the use of recurrent and time-convolutional neural network models to incorporate knowledge about the temporal structure within the data and the use of graph convolution and euclidean convolution to incorporate knowledge about spatial dependencies in the data. The second class includes methods to directly incorporate physical knowledge into the neural network in the form of differentiable layers that avoid violating physics constraints and thereby enable extrapolation in certain limits. 

\begin{figure}[h]
	\centering
	\includegraphics[width=0.95\linewidth]{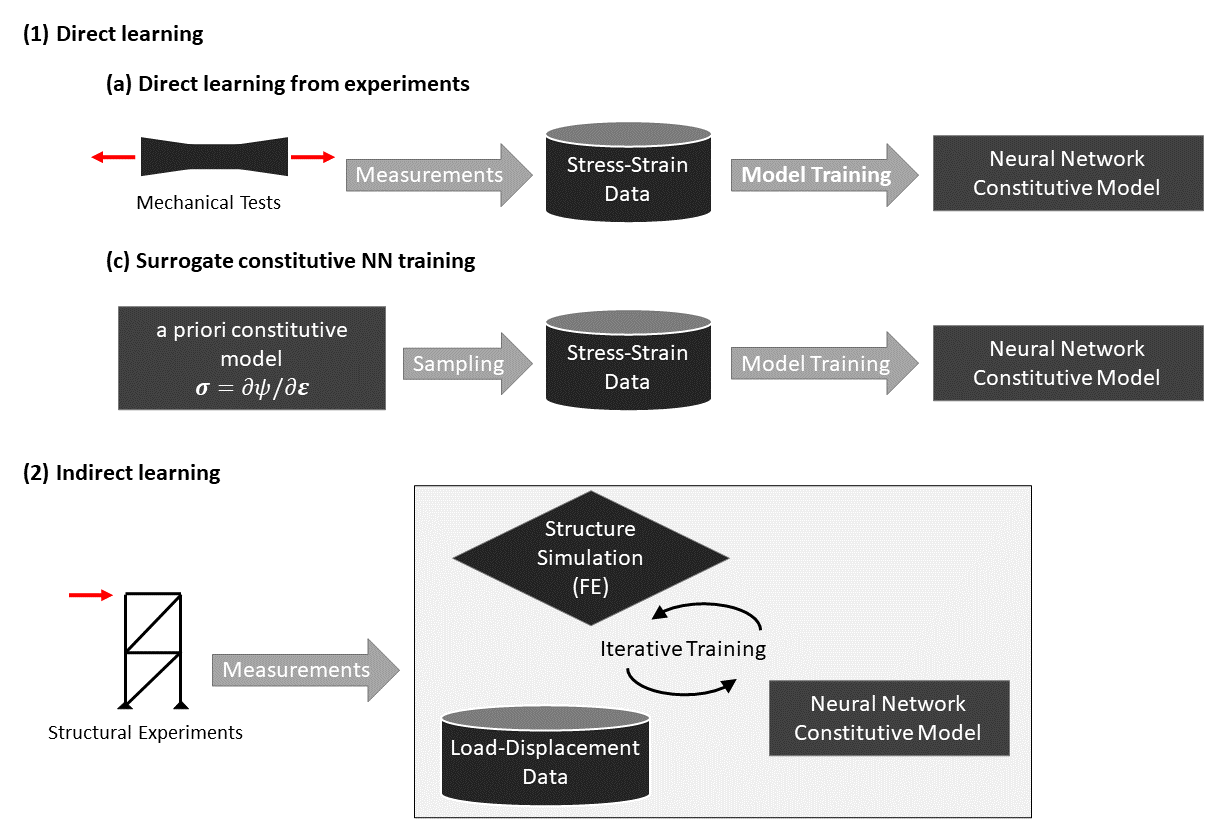}
	\caption{Applications of neural networks in constitutive modeling: (1) direct learning of constitutive neural networks based on (a) experimental stress-strain data and (b) a priori constitutive models, (2) indirect learning of constitutive neural networks in an iterative training scheme.}
	\label{fig:application_scenarios}
\end{figure}

Based on the application mode of neural networks-based constitutive modeling, two distinct use cases of learning constitutive relations with neural networks can be categorized, as illustrated in Figure \ref{fig:application_scenarios}. These are:
\begin{enumerate}[label=(\arabic*)]
	\item \textbf{Direct learning} is attributed to neural networks, which are trained directly on stress-strain data. Direct learning can be subdivided further into learning from experimental data and learning from synthetic data (constitutive surrogate modeling):
	\begin{enumerate}[label=(\alph*)]
	\item \textbf{Direct learning from global experimental stress-strain data} offers means to establish unidentified constitutive relations directly from the data. This approach is particularly valuable when insufficient knowledge about the specific material is available or when purely physical constitutive models are challenging to acquire. This approach has a long standing tradition and has been widely investigated since the early 90s \cite{ghaboussi1991knowledge}. However, a crucial prerequisite for direct learning is the availability of an adequate quantity of stress-strain data, commonly acquired from simple mechanical tests that yield one data sample per test run. In addition, these tests often assume stress-strain uniformity along a single direction, thereby limiting the data generation to one-dimensional measurements \cite{tao2021learning}.
	\item Learning \textbf{constitutive surrogates} is based on data sampled from reference classical constitutive models and aims to replace them to facilitate real time computations. Both direct learning from global experimental data and the learning of constitutive surrogates can be methodically categorized as standard supervised regression problems. However, the underlying motivation for these scenarios differ in that the former seeks to establish a constitutive relation, while the latter aims to replace a pre-existing, a priori given model in a computationally efficient framework. In addition, a neural network is classified as a constitutive surrogate if it is a function of a history variable that can be obtained only using a classical constitutive relation, and not from experiments. Although this method's ability to uncover novel material mechanisms is constrained by the physics-based used to generate the training data at a particular length scale. Utilizing this approach effectively to bridge multiple length scales holds the potential to discover mechanisms at a higher scale.
	\end{enumerate}
	\item To leverage the rich local constitutive information from structural tests, \textbf{indirect learning} utilizes numerical simulations in conjunction with experimental observation to train the constitutive neural network, in often an incremental manner. To facilitate the learning process, an error measure is established by comparing simulation results with experimental data. Gathering experimental data is, however, challenging and time-consuming, leading to a prevalent reliance on simulations alone as proof-of-concept in literature.
\end{enumerate}

\subsection{Paper Structure}
Our paper is structured as follows. In the remainder of this section we give a brief overview of machine learning and materials modeling background, which is assumed in the following sections. In Sections \ref{black_box} to \ref{grey_box}, we introduce and discuss the reviewed work. Wherein, we categorized the work according to the methodical spectrum from Figure \ref{fig:methods_overview}, where applications of classical FFNNs (category universal function approximation in Figure \ref{fig:methods_overview}) are reviewed in Section \ref{black_box}. In Sections \ref{advanced_recurrent} and \ref{advanced_conv} we discuss works that use advanced machine learning techniques that consider load history information and spatial information respectively. In Section \ref{grey_box}, we review work on physics-informed neural networks. Finally, in Section \ref{conclusions}, we give a summary, briefly discuss the current state of the field and give an outlook.

\subsection{Aspects of Machine Learning and Data Science}
In the following, we briefly describe the types of neural networks and dimensionality reduction techniques that are most important to methodically contextualize the literature reviewed throughout this paper. We refer to \cite{goodfellow2016deep} for a more comprehensive and in-depth overview of the field of deep learning.

\subsubsection{Neural Networks-Based Function Approximation}\label{background_nn}
\begin{figure}[h]
	\centering
	\includegraphics[width=0.7\linewidth]{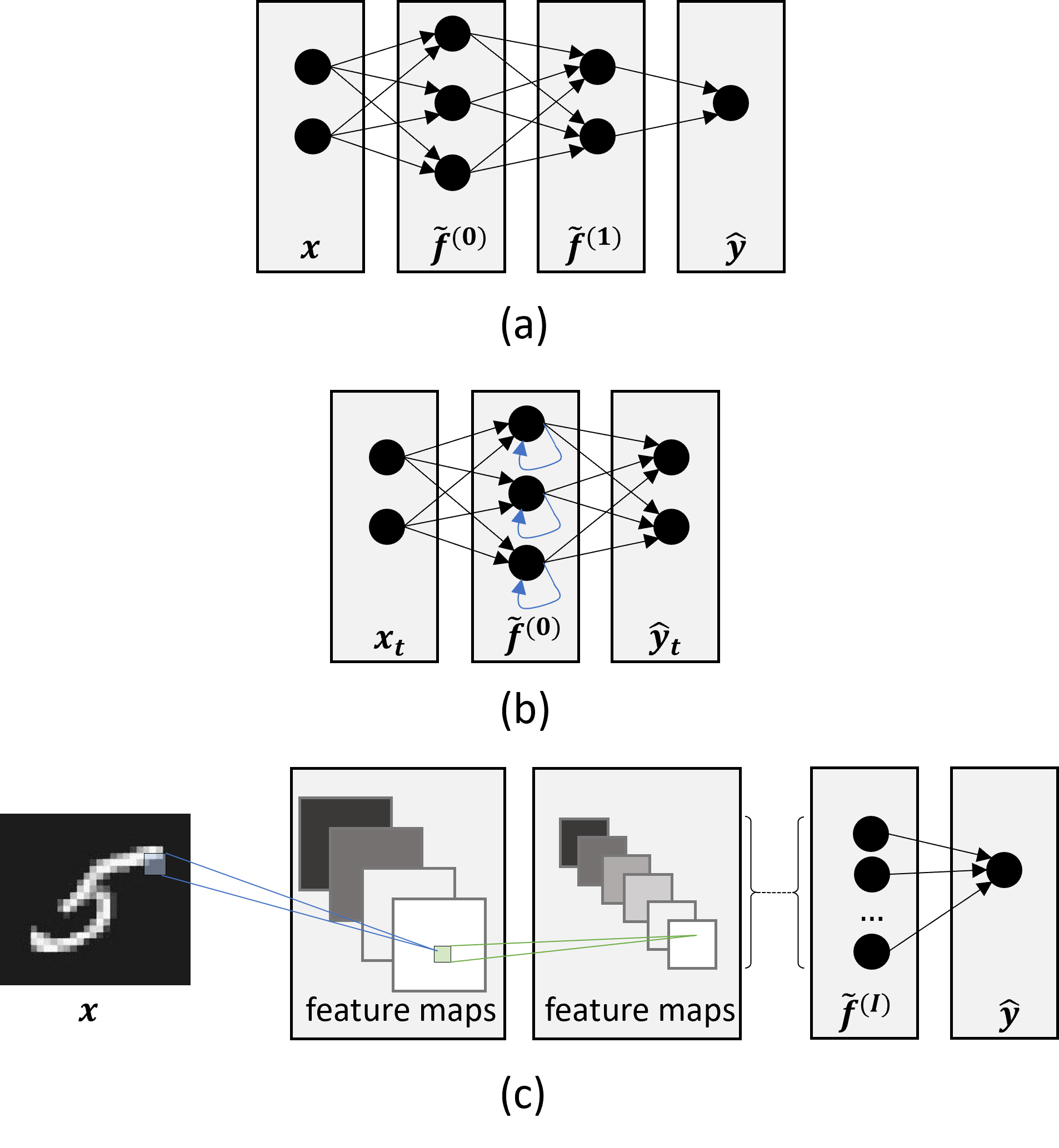}
	\caption{Basic architectures for the neural network types discussed throughout the paper. While the connections in FFNNs (a) form an acyclic graph, RNNs (b) contain cyclic connections (depicted in blue color), which cause the output of the respective neurons affecting the subsequent input of the same neurons. CNNs (c) are designed to process structured data, like images, time series or graphs.}
	\label{fig:nn-types}
\end{figure}

In this Section, we give an overview of basic neural network types used for constitutive modeling in the reviewed publications, including FFNNs, RNNs and CNNs. The outlined neural network types are depicted in Figure \ref{fig:nn-types}. FFNNs are general mappings

\begin{equation}
	\hat{\boldsymbol{y}}=\tilde{f}(\boldsymbol{x},\theta)
\end{equation}
with model parameters $\theta$, which can be trained for non-linear approximation of typically real-valued functions \begin{equation}
	\boldsymbol{y}=f(\boldsymbol{x}),
\end{equation}
with $\boldsymbol{x} \in\mathbb{R}^m$ and $\boldsymbol{y},\hat{\boldsymbol{y}}\in\mathbb{R}^n$. Gradient-based methods and the backpropagation algorithms are used to fit the parameters $\theta\in\mathbb{R}^l$ from sample data to approximate $f$. 

an FFNN consists of several consecutive neural layers
\begin{equation}
	\bar{\boldsymbol{y}}=\tilde{f}^{(i)}(\bar{\boldsymbol{x}}, \theta_i) 
\end{equation}
with parameters $\theta_i$. The information flow is directed from the input layer $\tilde{f}^{(0)}(\boldsymbol{x},\theta_0)$ through the hidden layers to the output layer $\tilde{f}^{(I)}$. The FFNN can then be seen as a nested function 
\begin{equation}
	\tilde{f}(\boldsymbol{x}, \theta)=\tilde{f}^{(I)}(\tilde{f}^{(I-1)}...(\tilde{f}^{(0)}(\boldsymbol{x},\theta_0)))
\end{equation}
with $\theta_i\subset\theta$ for $i\in [0,...,I]$. In the case of FFNNs, each layer ${f}^{(i)}$ consists of a set of processing units, so-called neurons, with a subset of $\theta_i$ as parameters. The amount of neurons per layer is referred to as the width of the layer, while the amount $I$ of layers in a neural network is referred to as its depth. The training of neural networks is typically conducted by mini-batch-wise application of a gradient descent based algorithm \cite{ruder2016overview}.

While FFNNs are generally applicable, more advanced neural network models take advantage of the data structure at hand to improve accuracy and data efficiency. Three important classes of such structured data are sequential data, image data, and graph data. Pertaining to sequential input data (like for predicting nonlinear microstructure evolution \cite{montes2021accelerating}), which is often present in materials modeling as the material behavior is typically history-dependent, recurrent or time-convolution neural networks can be used. RNNs \cite{rumelhart1986learning, elman1990finding} have special feedback connections, which makes them differ from the hierarchical structure of an FFNN. While the input layer of an RNN takes data from the current time-step, the recurrent connections enable information flow in between the time steps, and thereby the current output $\hat{\boldsymbol{y}}_t$ depends not only on the current input $\boldsymbol{x}_t$, but also on all previous inputs $\boldsymbol{x}_{t-i}$ for $i\in [0,t]$. 

For long sequences, these dependencies often lead to problems with the flow of the gradients through the model. During backpropagation, the gradients flow from the output neurons, where the loss is calculated based on $\hat{\boldsymbol{y}}_t$, through the unfolded recurrent connections to the input neurons. For longer sequences, this chain of calculations gets bigger and the gradients tend to vanish (tend to zero) or to explode (tend to infinity). While in the first case, the network cannot learn long-term dependencies, in the second case, the learning stability is often diminished. Due to this problem, initially proposed RNNs are considered impractical for complex tasks involving sequential data.
The long short-term memory network (LSTM) \cite{hochreiter1997long} is an attempt to solve the vanishing gradient problem based on learned input and output gates, which guard the flow of information. The gated recurrent unit (GRU) \cite{cho2014properties} is a simplified LSTM variant with a reduced amount of parameters. Both, GRUs and LSTMs, have been shown to often perform better than conventional RNNs in the case of small data sets, while there is no clear superiority of GRUs and LSTMs over the other \cite{chung2014empirical}. 

% Convolution
Another important type of deep neural network model is the CNN \cite{fukushima1988neocognitron, lecun2015deep}. In addition to the conventional fully connected layers, CNN models are composed of a range of specialized layers, with the convolutional layers being the most significant among them. In these layers, convolution kernels with trainable weight parameters efficiently learn abstractions and features of the structured input. Especially in the area of computer vision, CNNs led to radical progress in terms of accuracy and sample efficiency. In materials modeling, CNN models are often used in the context of multi-scale modeling, where image or graph data on the crystallographic scale are to be processed. Besides images, which are processed by two-dimensional convolution operations, one-dimensional convolution layers can be applied to sequential data \cite{cecen2018material, durmaz2021deep}. Graph convolution, as a generalization of the classical convolution, enables the application of convolutional layers on data that have no regular (Euclidean) structure but are instead structured in a more general form of a graph, see \cite{thomas2023materials} for an example in materials science.

\subsubsection{Dimensionality Reduction, Representations and Symmetries}\label{background_representations}

When working with results from experiments and simulations, often a huge amount of data is extracted, in which the relevant information is covered by a large amount of highly intercorrelating features. Moreover, material data, for example crystallographic texture data, often hold certain symmetries. In the following, we give a brief overview of methods to deal with these characteristics of data and to reduce the data dimensionality.

Dimensionality reduction techniques aim to embed high-dimensional data into a lower-dimensional embedding space, which is easier to process and interpret while preserving as much information as possible. The most prominent dimensionality reduction method is the principal component analysis (PCA), which projects the data linearly into a new coordinate system of lower dimension while preserving most of the variation within the data. In contrast, non-linear dimensionality reduction such as manifold learning methods (e.g. Isomap \cite{tenenbaum2000global}, multidimensional scaling \cite{cox2001multidimensional} or local linear embedding \cite{saul2003think}) are based on the assumption that the dimensionality of datasets is artificially inflated. Therefore, the data is extracted from a low dimensional manifold that is integrated within the high dimensional space of the data set \cite{cayton2005algorithms}.

A prominent approach to reduce the dimensionality of a data set nowadays is to use so-called autoencoder neural networks \cite{rumelhart1985learning}. This method has already shown to be useful for compressing microstructure information, see \cite{jung2020microstructure,iraki2021multi}. Such networks consist of an encoder and a decoder part. The encoder in-turn consists of fully connected layers only or a combination of convolutional layers with pooling and fully connected layers, depending on the input representation. In both cases, the output of each layer is usually of lower dimension than the input, and the aim of the encoder is to embed the data in a low-dimensional latent feature space, which is represented by a neural network layer, called the bottleneck layer. The decoder part consists again of fully connected layers which can be combined with deconvolutional layers and map the bottleneck layer to the network's output layer, which is of the same dimension as the encoder input. The encoder and decoder parts are trained simultaneously, by a loss that punishes the dissimilarity of the encoder input and the decoder output and thereby drives the autoencoder to encode as much of the data information as possible into the low dimensional bottleneck layer.

As in other domains, such as image processing, data in material sciences often holds symmetries. Mathematical operations on the data are often equivariant or invariant with respect to certain transformations. The efficiency of machine learning methods often relies on how these invariances and equivariances are dealt with. The success of CNNs on image data, for example, depends on the consideration of shift-invariance or equivariance by the application of a sliding convolution kernel, where the convolution parameters are learned independently of the location of the kernel. Recent research on deep learning through the lens of geometric symmetries can be found in the very active field of geometric deep learning \cite{bronstein2021geometric}. Another approach to consider the assumed symmetries within the data is the augmentation of training data, e.g. by duplicating samples and manually manipulating the input, which is state-of-the-art in machine learning.

\subsection{Aspects of Materials Modeling}
For a prolonged period, materials science has been dominated by empirical, model-based theoretical, and computational research. However, in the last two decades, it has been extended to incorporate (big) data-driven science \cite{agrawal2016perspective}. For now, material models have been developed mainly based on empirical investigations and theories derived from physics relations. In the second half of the last century, the field of computational mechanics evolved, which enabled model-based evaluations for example by conducting simulations on different length scales, cf. \cite{raabe1998computational}.

% Haupt Seite 153 ff
On a continuum scale, following the theory of continuum mechanics and thermodynamics, universal balance equations exist (the balance of mass, linear momentum, angular momentum, energy, and entropy), which form the basis to calculate unknown mechanical field quantities (for motion, temperature, etc.). However, for solving mechanical problems, these equations are underdetermined. To determine all unknowns, material-related constitutive relations (e.g. stress-strain relationships) are needed. These constitutive relations are not universal and have to be formulated depending on the material at hand and its individual deformation behavior (see for example \cite{callister2018materials} for an introduction to material behavior and deformation mechanisms).

In classic continuum mechanics, different theories of materials behavior have been developed that describe elastic and viscous fluids as well as elastic, visco-elastic, plastic and visco-plastic solids, see \cite{haupt2013continuum}. A fundamental characteristic of elastic material behavior is that the stress state depends only on the actual deformation state. Inelastic material behavior, in contrast, depends on the entire deformation history. In such material models, the deformation history is typically represented by internal variables, such as the accumulated plastic strain. 

For developing constitutive models, several approaches exist. Driven by experimental investigations, phenomenological material models describe phenomenons observed in experiments, such as the Hall-Petch relation describing the grain size dependent plastic yielding \cite{hansen2004hall}, the Armstrong-Frederick model describing the Bauschinger effect in plasticity, and rheological models describing linear visco-elastic material behavior on the macroscopic scale \cite{haupt2013continuum}.

Material models on different length scales can also be formulated in the framework of thermodynamics principles. In such models, the modeling perspective focuses on energy storage and release, as well as dissipation effects. Therefore, thermodynamic potentials must be introduced based on the physical understanding of the underlying mechanisms and experimental observations. In addition, dissipation phenomena are often addressed by using internal variables theory. If the second law of thermodynamics is fulfilled, the model is called thermodynamically consistent. Basic approaches for developing thermodynamically consistent material models can be found in \cite{haupt2013continuum}.

As an alternative to formulating macroscopic material models from scratch and calibrating them to experimental measurements, the material microstructure and its behavior can be modeled directly in a so-called representative volume element (RVE). By using this technique, the initial boundary value problem for representing the microstructural behavior is typically solved numerically. By virtual testing procedures, the deformation behavior can be analyzed on the basis of the Finite Element method (FEM) \cite{roters2010overview} or fast Fourier transform (FFT) \cite{eisenlohr2013spectral}. The advantage of using RVEs is that arbitrary load cases can be applied, which are often hard to apply in real experiments, cf. \cite{butz2019parameter, wessel2021new, wessel2022machine}. The approach to calibrate macroscopic material models to RVE data can be called virtual laboratory \cite{zhang2016virtual}. The virtual laboratory can also be used to model scale transitions in whole process chains, as is shown in \cite{butz2010modeling} on the example of dual-phase steel production.

Furthermore, RVE-based models can be used directly in Finite Element (FE) component simulations, which is typically called FE$^2$ \cite{smit1998prediction, feyel2000fe2}. FE$^2$ corresponds to the case where on both scales, the micro and the macro-scale, coupled FE simulations are performed to include the microstructural behavior in the simulated component. This in-turn allows for microstructure-driven design and optimization of manufacturing processes and components, cf. \cite{olson1997computational}. However, FE$^2$ is usually too time-consuming for the use in engineering. To overcome this issue, machine learning techniques can be applied, as will be presented in the following sections of this review. Instead of using FEM to determine the stress response of an RVE under external load, for example, supervised learning can be used to learn the stress responses for given deformation paths \cite{mozaffar2019deep}. The learned model forms an RVE surrogate model acting on the macroscopic scale, which is, in contrast to phenomenological models, sensitive to the underlying microstructure.

\section{Universal Function Approximators: Fully Connected Neural Networks for Learning Constitutive Relations}
\label{black_box}
In this section, we discuss work that uses fully connected neural networks to learn constitutive models, 
so-called constitutive neural networks  (also called neural constitutive laws \cite{pernot1999application} or neural network constitutive model \cite{shin2000self}. 
The section is structured into two subsections according to the application cases mentioned 
in \ref{methods_classification}. The use of FFNNs to directly learn the constitutive behavior of materials from global experimental stress-strain data 
as well as from synthetic simulation data is discussed in \ref{black_box_direct}, and work on indirect training schemes with the goal to learn constitutive neural 
networks from structural experiments is introduced and discussed in Section \ref{black_box_indirect}. The presented approaches are discussed briefly in Section \ref{sec:critique_universal_func}.

\subsection{Direct Learning}\label{black_box_direct}

\begin{figure}[h]
	\centering
	\includegraphics[width=0.72\linewidth]{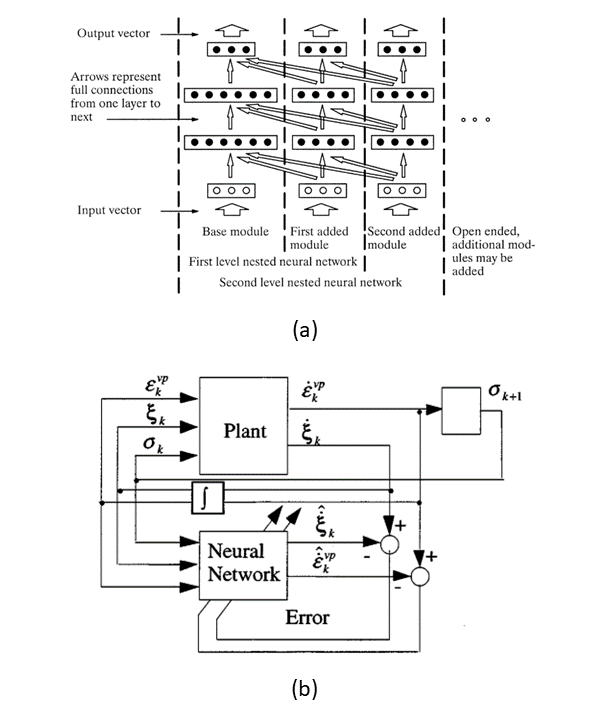}
	\caption{Early approaches to constitutive neural networks. (a) The nested adaptive neural network architecture introduced in \cite{ghaboussi1998new} Copyright\textsuperscript{\tiny\textcopyright} 1998, Elsevier. (b) Embedding graph of the implicit constitutive neural network introduced in \cite{furukawa1998implicit} Copyright\textsuperscript{\tiny\textcopyright} 1998, John Wiley and Sons.}
	\label{fig:early_approaches}
\end{figure}

Ghaboussi et al. proposed the use of FFNNs for constitutive modeling in a series of early works \cite{ghaboussi1991knowledge, ghaboussi1998new}. In \cite{ghaboussi1991knowledge}, the application of FFNNs on experimental data of concrete is proposed. The experimental data sets of biaxial monotonic loading from \cite{kupfer1969behavior} and uniaxial cyclic loading from \cite{sinha1964stress} is used for training and evaluation of FFNNs. For the monotonic biaxial loading case, both, stress-controlled and strain-controlled models, are put forth. In case of stress-controlled models, the neural network predicts the strain increment $\Delta\boldsymbol{\varepsilon}$ based on stress-strain states $\boldsymbol{\sigma}^{(t-i)},\boldsymbol{\varepsilon}^{(t-i)}$ sampled from previous time steps and the current stress increment $\Delta\boldsymbol{\sigma}$.
\begin{equation}
	(\Delta\varepsilon_1,\Delta\varepsilon_2)=f(\sigma_1,\sigma_2,\varepsilon_1, \varepsilon_2,\Delta\sigma_1,\Delta\sigma_2,\theta_\sigma).
\end{equation}
In strain-controlled models, increments of strain are provided as inputs and increments
of stresses are obtained as output, given by the relation
\begin{equation}
	(\Delta\sigma_1,\Delta\sigma_2)=f(\sigma_1,\sigma_2,\varepsilon_1, \varepsilon_2,\Delta\varepsilon_1,\Delta\varepsilon_2,\theta_\varepsilon),
\end{equation}
where, $\theta_\sigma$ and $\theta_\varepsilon$ are the neural network parameters. The strain-controlled model has the advantage of being directly usable in FE simulations. For uniaxial cyclic loading, a two-layer constitutive neural network which takes in two previous points on the stress-strain curve, in addition to the current point is proposed. The history of the stress-strain curve is important to capture the cyclic behavior. In the results, the authors show the ability of the neural networks to generalize to unseen proportional and non proportional stress paths for monotonic loading, but it falls short for unseen low stress cyclic loading path.

As a methodic extension to previous work, Ghaboussi et al. \cite{ghaboussi1998new} proposed the nested adaptive neural network (NANN). The NANN is an FFNN, with a special structure of successively learned nested stages. The NANN, as depicted in Figure \ref{fig:early_approaches} (a), consists of a base module, which corresponds to a strain-controlled constitutive FFNN
\begin{equation}
	\Delta\sigma^{(t)}=f(\Delta\varepsilon^{(t)},\sigma^{(t)},\varepsilon^{(t)},\theta).
\end{equation}	
The basis module is extended by gradually trained additional modules
\begin{equation}
	\Delta\sigma^{(t)}=f(\sigma^{(t-i)},\varepsilon^{(t-i)},\theta)
\end{equation}
of path-dependent deformation data for discrete time steps $(t-i)$ to form higher level NANNs, where $\theta$ are the associated neural network parameters. The method is tested on experimental data of triaxial compression test of sand, for which the authors show how the prediction quality increases depending on the NANN level. Although the proposed architecture enables training constitutive neural networks from data describing variable length paths, the model complexity (i.e. the number of trainable parameters) grows linearly with the maximum level of the NANN and the method is therefore limited to short stress-strain paths. Due to the variable length input, the proposed method can be seen as a precursor of the later proposed applications of RNNs, which are discussed in subsequent Section \ref{advanced_recurrent}.

Unlike in \cite{ghaboussi1991knowledge} and \cite{ghaboussi1998new}, where the internal material state is assumed to be captured sufficiently by discrete samples from the deformation history, Furukawa et al. \cite{furukawa1998implicit} proposed to represent the material state in a state space model and to train a so-called implicit constitutive neural networks. Inspired by control theory, the implicit constitutive neural network acts as a surrogate of a dynamical system
\begin{equation}
	\dot{\boldsymbol{x}}=f(\boldsymbol{x}, \boldsymbol{u}). 
\end{equation}
Where, for visco-plastic material models, Furukawa et al. defines the state
\begin{equation}
	\boldsymbol{x}=(\varepsilon_\mathrm{vp}, \zeta)
\end{equation}
as a combination of the visco-plastic strain $\varepsilon_\mathrm{vp}$ and internal variables $\zeta$, such as the back stress and drag stress. The stress plays the role of the control input $\boldsymbol{u}$. As shown in Figure \ref{fig:early_approaches} (b), the neural network models the influence of stress on the internal variables and strain rate, while the stress for the next time step is derived analytically. The proposed framework is instantiated for visco-plasticity and tested on pseudo-experimental uniaxial cyclic-loading data, which are sampled from the Chaboche's constitutive model \cite{chaboche1989constitutive}, and on experimental data of steel on elevated temperature under the assumption of fixed strain rates and known elastic responses. Although comparable results of the implicit constitutive neural network to that of 1D Chaboche's constitutive model have been shown, its validity is not proven for a higher dimensional case, and for non proportional complex stress states.

Further, Lefik and Schrefler \cite{lefik2003artificial} proposed history-dependent constitutive neural networks to model material behavior of super-conducting fibers under biaxial loading with hysteresis. Several points on obtained stress-strain curves are used to describe the material state. For training, data from numerical simulations are used, for which a rotation data augmentation technique is applied to cover invariances within the data and to preserve objectivity of the constitutive model. The rotation operation is performed without additional constraints in the cross section of the fiber, where the material model behaves isotropic. Therefore, common rotation matrices $\boldsymbol{R}$ were applied to the relevant quantities, like for example to the stress tensor
\begin{equation}
\boldsymbol{\sigma}_\mathrm{rot} = \boldsymbol{R}^\top \boldsymbol{\sigma} \boldsymbol{R}, ~~~
\mathrm{with} ~ \boldsymbol{R}^\top\boldsymbol{R} = \boldsymbol{1}.
\end{equation}
The elasto-plastic constitutive neural network is applied to reproduce the behavior simulated in one dimension, as well as the homogenized behavior of super-conducting fibers from a two-dimensional simulation. The former results are compared to experimental results. 

Al-Haik et al. \cite{al2006prediction} proposed the use of FFNNs to predict the relaxation of polymeric matrix composites depending on constant strain and temperature conditions. The model maps from strain, temperature, and process time to the relaxation stress and is trained on data from stress relaxation tests of a carbon fiber epoxy composite. The authors highlight that their model is in general more accurate than an explicit conventional visco-elastic model, in particular for temperatures near the glass transition temperature. 

In more recent work, FFNNs are applied to learn from direct stress-strain data from various metallic materials \cite{gorji2019towards, du2020modeling} and polypropylene \cite{jordan2020neural}. By Gorji et al. \cite{gorji2019towards}, FFNN models are learned on pseudo-experimental data of sheet metal generated by using (a) the Zerilli-Armstrong model for temperature and strain rate-dependent hardening and (b) a J2 plasticity model used to simulate biaxial monotonic loading. The results show a good agreement of the FFNN response with responses of the J2 model for uniaxial notched tensile tests. Furthermore, Jordan et al. \cite{jordan2020neural} used FFNNs with Bayesian regularization \cite{mackay1992bayesian} to model the temperature and strain rate-dependent behavior of polypropylene based on experimental uniaxial loading data with varying temperature and strain rates. The FFNN learns the mapping from the viscous strain, viscous strain rate, and temperature to true stress. The comparison with a state-of-the-art conventional thermo-elastic visco-plastic model, which is calibrated to the experimental data, reveals a predominance of the FFNN in terms of accuracy and computational performance.

Also, for the direct learning setting, du Bos et al. \cite{du2020modeling} proposed and evaluated an alternative procedure, in which neural networks are trained to map from strain paths to corresponding stresses. The estimated stress paths are interpolated afterwards to reconstruct the stress-strain curve. In difference to the majority of the methods referenced above, which perform predictions of the material behavior at each time step based on current and previous conditions, the proposed model aims to predict the global stress-strain relation for a given strain path in one pass. To reach a good model performance with a reduced input and output space, the sampling technique is optimized by minimizing the interpolation error. The method is evaluated on strain paths from reverse loading cases applied to isotropic and rate-independent elasto-plastic solids, for which it is shown to reach an acceptable prediction quality in the low data regime (in particular, 100 samples were used). Model training and evaluation was carried out on pseudo-experimental data. 

\begin{figure}[h]
	\centering
	\includegraphics[width=0.4\linewidth]{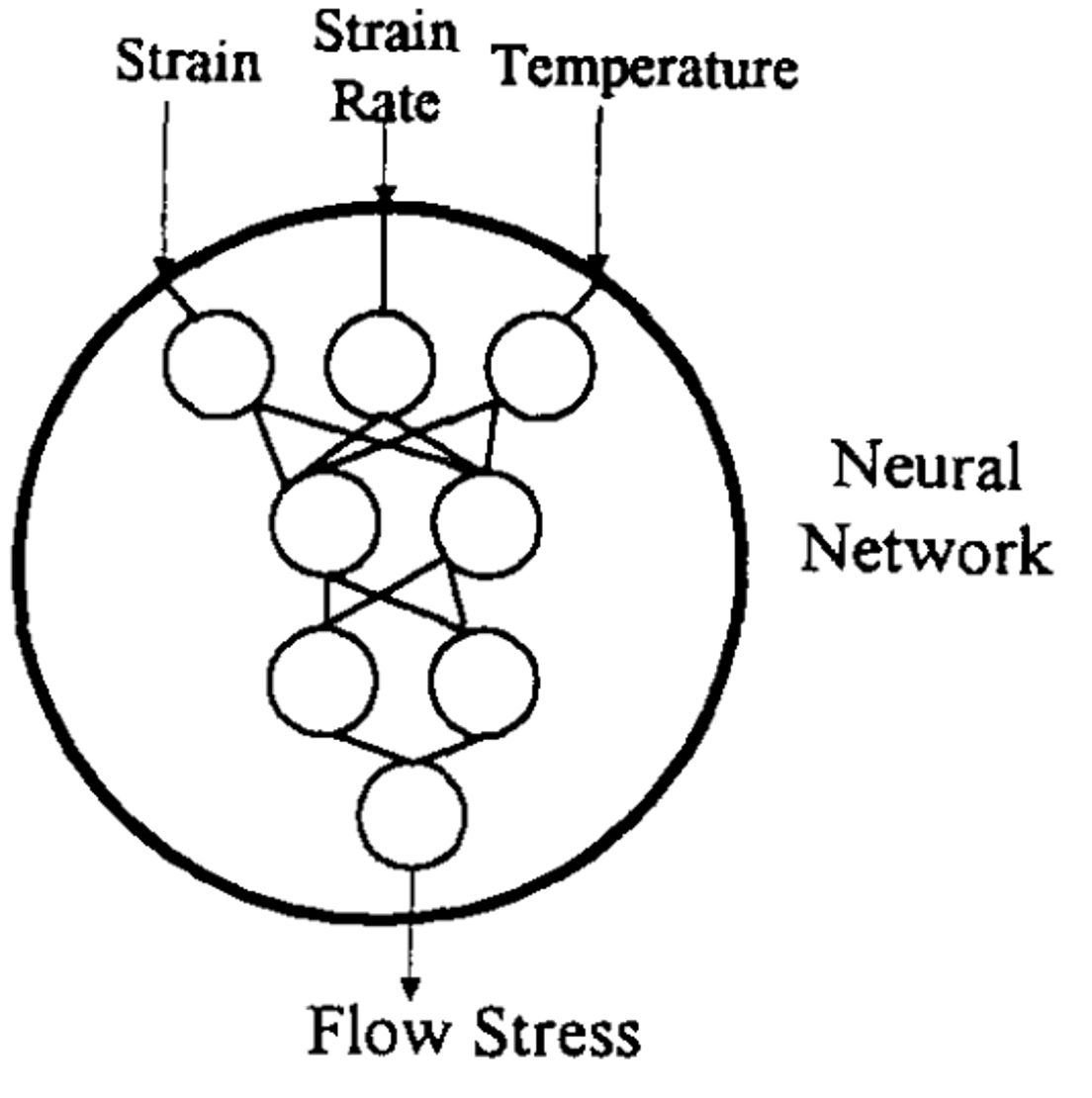}
	\caption{Early general network structure for flow stress prediction introduced in \cite{rao1995neural} Copyright\textsuperscript{\tiny\textcopyright} 1995, Elsevier}
	\label{fig:high_temp}
\end{figure}

When it comes to metallic materials, neural networks were used quite early for directly learning the hardening model from experimental high-temperature deformation data. In hot forming processes, many interrelated and non-linear hardening and softening phenomena such as work hardening, dynamic recovery, and recrystallization determine the constitutive flow behavior. The underlying mechanisms are significantly affected by temperature and strain rate. In this situation, conventional models with empirically fitted constants are often inaccurate or the applicable temperature and strain rate range is very limited \cite{lin2008application,ji2011comparative}. The parameters that define thermomechanical treatment also play a significant role for the microstructure and mechanical features of the hot formed product \cite{sani2018modeling}. Therefore, accurate hardening models are an important requirement for process optimization and process analysis \cite{rezaei2021hot, kumar2021construction, li1998approach}. In several works, neural networks are proposed as black box models for modeling the flow behavior of various types of steels \cite{rao1995neural,hodgson1999prediction,mandal2006constitutive,lin2008application,ji2011comparative, li2012comparative,han2013comparative,gupta2013development,kumar2021construction}, 
aluminium alloys \cite{chun1999using,bobbili2015prediction,ashtiani2016comparative,li2019constitutive}, 
titanium alloys \cite{li1998approach,sun2010development}, 
magnesium alloys \cite{sani2018modeling} 
and pure aluminium \cite{rezaei2021hot}. 

Already in the mid 90s, Rao and Prasad \cite{rao1995neural} trained a constitutive neural network on experimental data of medium carbon steel deformations under a variation of constant strain rates and temperatures. As depicted in Figure \ref{fig:high_temp}, the neural network learns to map directly from the strain and the respective processing conditions to the flow stress. The model is shown to outperform a semi-empirical constitutive model. Similarly, Li et al. \cite{li1998approach} trained and evaluated a constitutive neural network to predict the flow stress of a titanium alloy at elevated temperatures based on experimental data.
Chun et al. \cite{chun1999using} also proposed constitutive neural networks to predict the flow stress during hot compression and rolling of aluminium alloys.
Hodgson et al. \cite{hodgson1999prediction} proposed a constitutive neural network that includes additional input parameters such as the work hardening and evaluated it on experimental hot torsion data of 304 stainless steel. It is shown that the model with additional parameters clearly outperforms both, a constitutive neural network without the additional parameters and a conventional phenomenological model.

Moreover, Mandal et al. \cite{mandal2006constitutive} proposed the use of a neural network to predict the flow stress of austenitic stainless steels from hot compression test data. Unlike the above mentioned publications, in \cite{mandal2006constitutive}, the constitutive neural network  also takes into account the alloy composition and can thereby be used to optimize the composition in addition to the hot forming parameters. Data from various grades of austenitic stainless steels are used and the composition is represented by the normalized amount of the most common elements. To be able to generalize across the composition, the amount of samples used for training and testing lies above $2000$, which is about one magnitude above the amount of data used for training in the above mentioned works dealing with flow stress prediction. In addition to a quantitative evaluation of the model, a sensitivity analysis of the neural network was carried out to estimate the relative importance of the input parameters, especially the chemical composition.

Furthermore, Lin et al. \cite{lin2008application} trained a neural network to predict the constitutive flow behavior of 42CrMo steel on data from hot compression tests. Sun et al. \cite{sun2010development} also trained a constitutive neural network on data of compression tests but of a titanium alloy. The model predictions show good agreement with the experimental data and the model is shown to clearly outperform a conventionally used hyperbolic sine model. Ji et al. \cite{ji2011comparative} trained a similar constitutive neural network on isothermal hot compression data of Aermet100 steel. The trained neural network is compared with a conventional Arrhenius-type constitutive model with strain compensation, which it outperforms in terms of accuracy. The measured mean absolute percentage error (MAPE, describing the mean over the relative regression errors) achieved by the conventional model is $7.62$ while it is $2.58$ for the constitutive neural network. At this point, however, we want to remark that while MAPE is a popular metric in studies of high-temperature deformation neural networks, it is known to be biased when used for model comparison as it assigns higher penalties to negative errors than to positive errors \cite{tofallis2015better}. Therefore, and due to issues with close-to-zero values MAPE is not commonly used in typical machine learning tasks and regression analysis nowadays. 

Anyhow, regarding the learning of the behavior of Aermet100 steel, in \cite{ji2011comparative}, it is shown that, in contrast to a trained constitutive neural network, the conventional model performance collapses in the instability regimes, where the physical mechanisms differ from the ones in the stable regime. The constitutive neural network manages to accurately predict the constitutive flow behavior over the whole experimental range of temperatures and strain rates. 
In a similar manner, in \cite{li2012comparative} the constitutive neural network approach is compared to a modified Zerilli-Armstrong and a strain-compensated Arrhenius-type model on hot compression experiments of T24 steel. Again, the result of the comparative study is that the constitutive neural network outperforms the conventional approaches (with a MAPE of $0.45$ for the constitutive neural network, $2.72$ for the Arrhenius-type model and $5.22$ for the Zerilli-Armstrong model). However, it is pointed out that this comes at the cost of interpretability and the requirement of high-quality data. 

Bobbili et al. \cite{bobbili2015prediction} trained a constitutive neural network to model the flow behavior under a high strain rate of 7017 aluminum alloy on data from the split Hopkinson pressure bar test \cite{hopkinson1914x}. In contrast to the majority of the publications in this section (where compression tests with strain rates up to 50 $s^{-1}$ are considered), in \cite{bobbili2015prediction}, the experimental data covers strain rates between 1500 $s^{-1}$ and 4500 $s^{-1}$. The trained constitutive neural network is compared to a Johnson-Cook model \cite{johnson1983constitutive}, and it is shown that the constitutive neural network outperforms the conventional baseline (with a MAPE of $2.58$ for the constitutive neural network and $10.62$ for the Johnson-Cook model). 

More recent work on hot deformation constitutive neural networks include \cite{sani2018modeling, li2019constitutive, rezaei2021hot, kumar2021construction}. 
Sani et al. \cite{sani2018modeling} proposed a neural networks-based approach for modeling the behavior of cast magnesium (Mg-Al-Ca) alloy and compare it with a conventional hyperbolic sine-based model on hot compression experimental results. It is shown that the hyperbolic sine functions accurately predict the stress-strain curves for cases in the high temperature and low strain regime, but are biased for cases with low temperature and high strains, where twinning effects play an important role. The trained neural network on the other hand is shown to accurately predict the relationship over the whole strain-temperature range. 
Li et al. \cite{li2019constitutive} compared the neural network approach with a phenomenological Arrhenius-type model and a physics-based model for work hardening and dynamic recovery on data from hot compression tests of 6082 aluminum alloy. As in the already mentioned comparative studies, the neural networks-based approach outperforms the conventional approaches according to the statistical metrics used, while the conventional methods have advantages in terms of model interpretability.

Moreover, Rezaei Ashtiani and Shayanpoor \cite{rezaei2021hot} proposed to use the initial grain size as an additional feature, besides strain, strain rate, and temperature, to predict the aluminum flow behavior under hot working conditions. Neural networks are trained based on data from isothermal hot compression tests of AA1070 specimens with varying initial grain sizes in the range $50\mu m-450\mu m$. Within the application study, the trained neural networks are used to determine processing maps and stable process regions depending on the initial grain size. While analyzing the relative sensitivity of the input parameters to the outputs, it is shown that temperature and initial grain size are most significant for the learned mapping, which is in accordance to well-known experimental observations, e.g. the Hall-Petch relation. Of course, this outcome mainly confirms that these effects are represented by the underlying data. As in \cite{rezaei2021hot}, Kumar et al. \cite{kumar2021construction} trained a neural network model and used it to create various processing maps, such as strain rate sensitivity maps, for 9CR-1Mo steel. Particularly, flow stress neural networks are proposed and compared to a linear interpolation stress correction scheme. In the results, the predominance of the neural network model is shown. 

In general, the approaches to directly learn from global experimental data heavily rely on idealized laboratory experiments, where deformation is kept homogeneous, and loading paths are nearly proportional. However, at the component level, deformations are heterogeneous, and the stress state and strain path often fall outside the experimental range. While the neural networks demonstrate promising results and occasionally outperform classical constitutive relations, their applicability beyond the training domain remains unproven. This is where the classical constitutive relation's advantage lies, as it can extrapolate predictions effectively. It's worth noting that a trained constitutive neural network has rarely been applied to predict at the component scale. Including the established constitutive relationship would help mitigate this issue, in addition to accelerating the current modeling strategies. This however comes at the cost of loss of universality of the neural network, as this is constrained by the constitutive equations. In the following, we review works on learning constitutive surrogates with the help of classical constitutive equations.

For building constitutive surrogates, neural networks are trained on data sampled from simulation results by using a reference constitutive model. The neural network surrogate model which is thereby obtained is used to replace the reference model in future simulations with the motivation to accelerate the simulation. 
Early works that use FFNNs to act as surrogate constitutive models includes \cite{hashash2004numerical, jung2006neural, yun2008new}. In recent years, some works propose and discuss FFNNs as surrogate constitutive models \cite{bessa2017framework,stoffel2018artificial, huang2020machine, zhang2020using, jang2021machine}. This revival of the usage of neural networks for constitutive modelling was, however, initialized by the work of Le et al. in 2015 \cite{le2015computational}. Therein, an FFNN-based homogenization approach was proposed for hyper-elastic materials.

\begin{figure}
	\centering
	\includegraphics[width=0.75\linewidth]{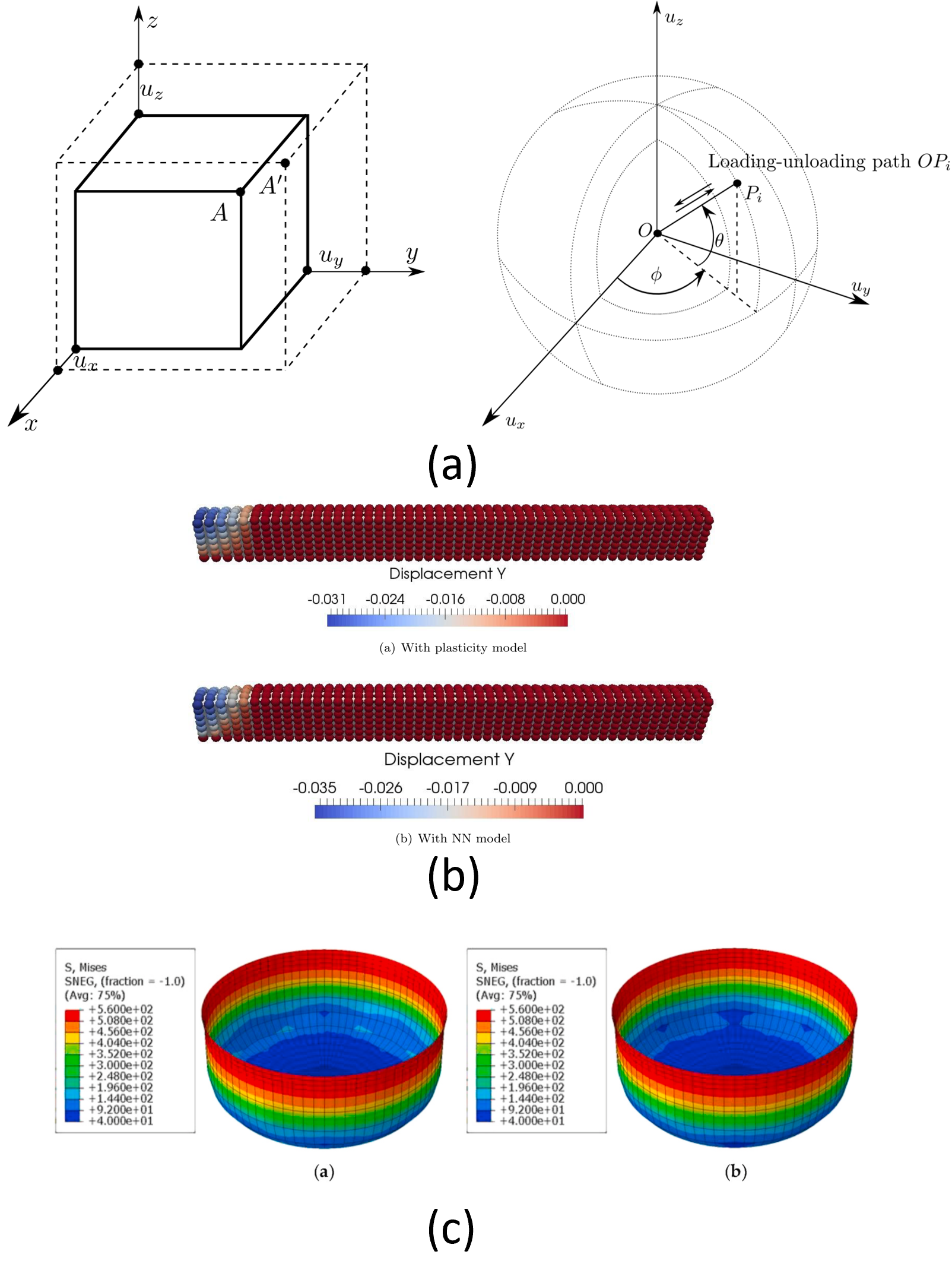}
	\caption{(a) Training data generation strategy and (b) exemplary results, utilizing the general constitutive surrogate described in \cite{huang2020machine} in an FE simulation of a tensile bar  Copyright\textsuperscript{\tiny\textcopyright} 2020, Elsevier. (c) Comparison of constitutive neural networks-based FE simulation results of a cup deep drawing process with the results of the original analysis based on a J2 plasticity model, from \cite{jang2021machine} Copyright\textsuperscript{\tiny\textcopyright} 2021, Elsevier.}
	\label{fig:surrogate_constitutive}
\end{figure}

In some of the previously discussed works in indirect learning, the integration of constitutive neural networks into FE simulations via user-defined material subroutines was utilized \cite{ghaboussi1998autoprogressive, sidarta1998constitutive}. Hashash et al. \cite{hashash2004numerical} addressed numerical implementation issues in this context. Therein, an explicit formulation of the material stiffness matrix is proposed, which is derived from constitutive neural networks and leads to a more efficient convergence behavior of FE simulations incorporating constitutive neural networks. However, Yang et al. \cite{yang2019derivation} pointed out, among other scenarios, that loading–unloading sequences are not considered in training the model proposed in \cite{hashash2004numerical}.

Jung and Ghaboussi \cite{jung2006neural} trained a rate-dependent isotropic constitutive FFNN, 
which maps from strains, stresses, and strain rates of the current and previous discrete time step and the previous rate of stresses to the current rate of stresses to model visco-elastic behavior. The network is trained from synthetic creep and relaxation test data, sampled for various step sizes, and then evaluated on data from concrete beams. It is pointed out that the approach performs well with variable time steps, in opposition to previous rate-dependent constitutive neural networks (e.g. \cite{furukawa1998implicit}). 
Yun et al. \cite{yun2008new} proposed the use of various internal material state variables in combination with the stress-strain information of the last time step as additional neural network inputs to predict the cyclic hysteresis behavior of materials. The proposed neural network is trained on simulated cyclic loading data of concrete and steel parts, implemented into the FE material model, and tested on structural simulations. The model showed superior prediction performance compared to earlier developed neural network constitutive models. The approach was evaluated on various uniaxial cyclic experiments and it is pointed out that the model showed the capability to learn post-limit material behavior, e.g. buckling, tearing, and yielding. Though, the influence of varying time step increments on the neural network prediction is not addressed.

A study on the performance of FFNNs is presented by Bessa et al. \cite{bessa2017framework} at the example of learning the material behavior of a 2D hyper-elastic composite. Particularly, this study investigates both, learning the material behavior on the basis of a constant RVE (including a comparison with the performance of a Gaussian process model \cite{rasmussenGP}) and for varying microstructure features. Moreover, this publication highlights the importance of the design of experiments, the importance of modeling uncertainty and the benefits of such machine learning-based surrogate models. 
In another work, Stoffel et al. \cite{stoffel2018artificial} compared two approaches to predict the highly dynamic behavior of shock-wave loaded plates. Both approaches involve the use of FFNNs. While in the first approach a (structural) FFNN is trained on experimental data to predict the structural deformation directly, in the second approach a constitutive neural network that aims to model the visco-plastic behavior is trained on simulation data and is implemented into FE code. The proposed constitutive surrogate maps from stress, backstress, and plastic strain tensor components to the plastic strain rate and backstress rate tensor components and eliminates the need for an iterative solution of the constitutive behavior. The direct comparison shows that the results of the neural networks-based constitutive surrogate are much more accurate than the results of the structural FFNN and reduce the computational effort of the FE simulation. In contrast, the effort for implementing the structural FFNN is much lower as it can be trained on experimental data only, and does not require a numerical simulation model.

Huang et al. \cite{huang2020machine} proposed a combination of the proper orthogonal decomposition and per-component constitutive neural networks to model hyper-elasticity and plasticity. The constitutive neural networks are trained on simulation results of 2D and 3D unit cell data with the accumulated absolute strain as a history variable to represent the material state. In the 3D case, data is sampled from a homogeneous cubic specimen under triaxial loading-unloading conditions, where the end-point $A'$ of each loading path is sampled randomly from a unit sphere as depicted in Figure \ref{fig:surrogate_constitutive} (a). To highlight the generalization abilities of the trained neural networks, the models are applied to various 2D and 3D structural FE simulations by using the automatic differenciation and derivation toolbox AceGen \cite{korelc2016automation} to derive the tangent matrix. As shown for a 3D bar necking example in Figure \ref{fig:surrogate_constitutive} (b), the results of these simulations are in good overall agreement with the results obtained when using the reference constitutive model.

Furthermore, Zhang and Mohr \cite{zhang2020using} proposed a constitutive neural network for von Mises plasticity with isotropic hardening. The neural network maps from the current stress and plastic work to the elasto-plastic tangent matrix and the Young's modulus without making a priori assumptions about the yield surface, flow rule, or hardening law and is combined with modified algorithms for uniaxial loading and plane stress loading. The approach is shown to be able to sufficiently reproduce predictions of a J2 plasticity model, including large deformation responses for complicated multi-axial loading-unloading paths. However, the transition between elastic and plastic domain had to be artificially smoothed, which can lead to over or underestimation of initial yield stress. 
Also, Jang et al. \cite{jang2021machine} proposed a neural networks-based surrogate of a J2 constitutive model, in which linear elastic loading and unloading is covered by a conventional physics-based model, while a nonlinear plastic correction is covered by the neural network. Due to this decoupling, the neural network can be trained efficiently based on one-element simulations and can be applied to a wide range of simulations. The trained model is verified on single-element simulations and tensile simulations of a dog-bone structure. Finally, a cup drawing simulation, depicted in Figure \ref{fig:surrogate_constitutive} (c), is conducted based on the neural network model.

The usage of FFNNs as surrogates for conventional constitutive models is often accompanied by the need for a high amount of data and does not always lead to the aimed acceleration of the simulation speed. In their study, Zhang and Mohr \cite{zhang2020using} showed that ten to hundred thousand data points have to be sampled for an accurate J2 surrogate model for strains up to $20\%$. In the 3D case described above, Huang et. al \cite{huang2020machine} train a model on samples from 8100 loading paths and emphasize that only loading is considered in the 3D case since the amount of data that would be needed to include unloading is too high. When it comes to simulation acceleration, Jang et al. \cite{jang2021machine} reported a speed increase of $11\%$ when using the neural network-based constitutive surrogate in comparison to the reference model. In \cite{huang2020machine}, even a decrease in computational performance is reported. 
Stoffel et al. \cite{stoffel2018artificial} reported a halving of the simulation time due to the proposed integration scheme. However, in computationally more complex scenarios, such as plasticity calculations on the microstructure level of polycrystals, constitutive neural networks are reported to provide a massive acceleration of simulations \cite{ali2019application,mianroodi2021teaching}.

\begin{figure}
	\centering
	\includegraphics[width=0.65\linewidth]{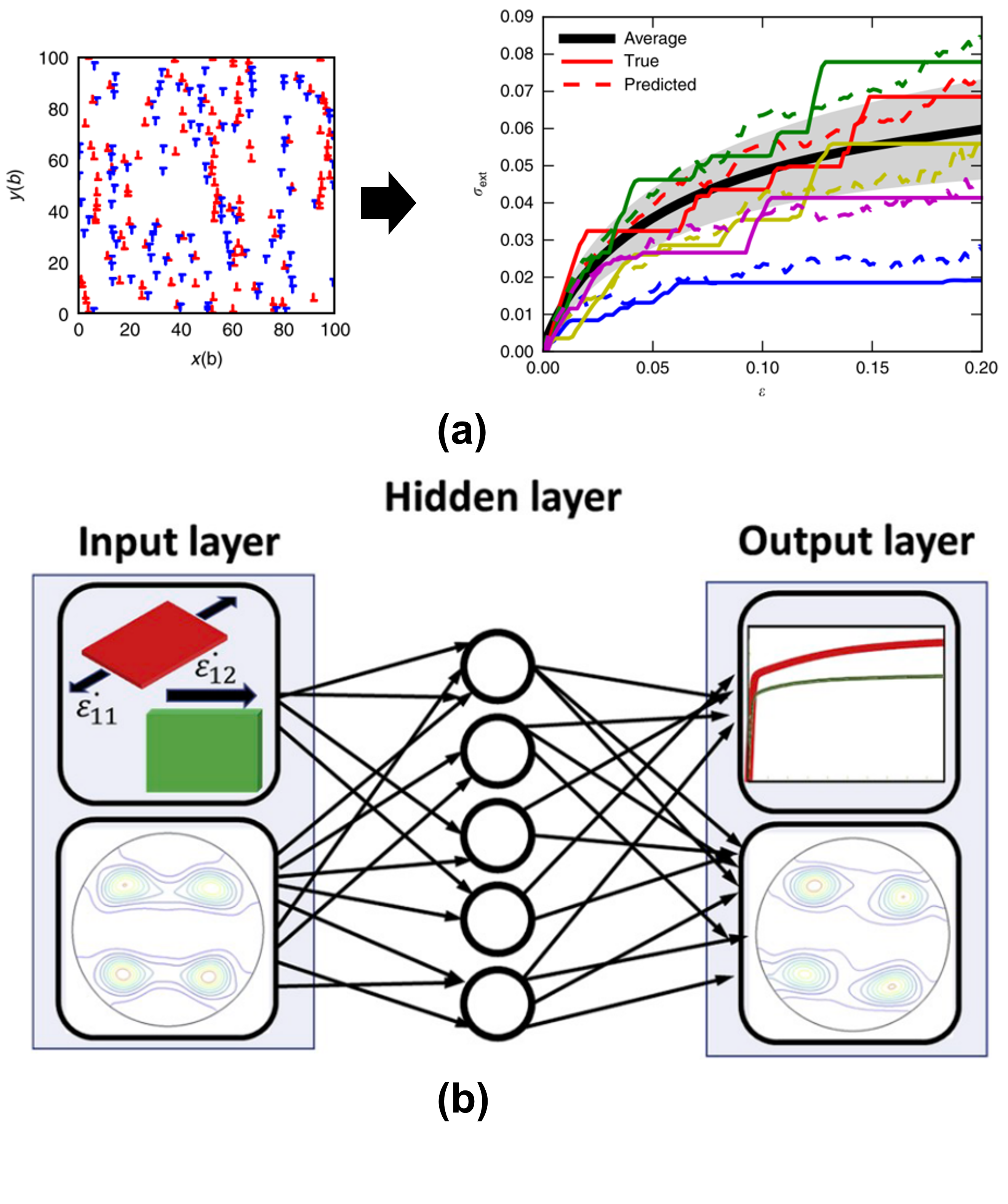}
	\caption{(a) The mapping (learned in \cite{salmenjoki2018machine}) from the dislocation configuration of a crystalline solid to the staircase-like stress-strain curve (reassembled from \cite{salmenjoki2018machine}, CC-BY). (b) Plot of the crystal constitutive neural network mapping introduced in \cite{ali2019application}, which learns the stress-strain relation together with the crystal evolution model Copyright\textsuperscript{\tiny\textcopyright} 2019, Elsevier.
	}
	\label{fig:blackbox_crystals}
\end{figure}

On the micro-scale of polycrystalline materials, FFNNs are applied in the context of constitutive modeling also. 
Salmenjoki et al. \cite{salmenjoki2018machine} proposed neural networks to learn a model of the behavior of micro-scale crystalline structures from two-dimensional discrete dislocation dynamics simulations. The work considers solids plastically deformed in a sequence of strain bursts, forming a staircase-like stress-strain curve, as shown on the right of Figure \ref{fig:blackbox_crystals} (a). The neural networks are trained to predict these curves from features originating from the initial dislocation configuration of the crystalline solid, which is depicted on the left of Figure \ref{fig:blackbox_crystals} (a), where red and blue symbols represent positive and negative Burgers vectors, respectively. Numerical experiments are conducted to quantify the approximation quality in various settings. The quality is found to be highly affected by the system size and, according to the authors, is surprisingly good for large-strain deformation dynamics. 

Furthermore, Ali et al. \cite{ali2019application} proposed FFNNs as fast surrogate models for the computationally expensive crystal plasticity Finite Element method (CPFEM), where the homogenized response of the RVE is of interest. As depicted in \ref{fig:blackbox_crystals} (b), the neural networks learn the stress-strain relation together with the texture evolution model. The neural networks are trained and evaluated based on data from experimentally validated rate-dependent CPFEM simulations of single crystals and polycrystal aluminum alloy AA6063-T6 under uniaxial tension and shear. The results show a good agreement with the CPFEM results and an immense decrease in calculation time (of up to $99.9\%$).

Ling et al. \cite{ling2016machine} compared two neural network approaches for constitutive modeling in the case of known symmetry and invariance properties. While the first approach is based on an augmentation of the data set to exploit the knowledge about the symmetries, the second approach is based on a proposed invariant representation of the input data. Besides the application of the approach to the modeling of a turbulent flow, the proposed methods are evaluated on a crystal-elasticity case study, with cubic crystal symmetry. 
The approach described in  \cite{ling2016machine} is generalize by Jones et al. \cite{jones2018machine} to embed further constraints and invariances for stress and plastic flow and show that this can reduce the amount of training required. 

\subsection{Indirect Learning}\label{black_box_indirect}

Learning constitutive neural networks directly from global experimental stress-strain data is a straightforward and easy-to-implement way towards a machine learning constitutive model, if some preconditions are fulfilled. To create a data set that is sufficient for neural network training, the constitutive relationships of interest have to be identifiable in sufficient quality, and a sufficient quantity of experiments to measure the data must be carried out. As stress fields in structural components are typically not measurable, training relies on the global material response measured. This means that each sample of the training set corresponds to one experiment, conducted usually under the assumption of uniform loading and homogeneous material behavior. Consequently, the direct learning approach is often restricted to simplified scenarios, such as the modeling of uniaxial and biaxial loading or the modeling of hardening behavior only. Compared to such tests, where each experiment is supposed to produce one stress-strain sample, results from structural experiments are rich in implicit constitutive information. The main goal of indirect learning of constitutive neural networks is to utilize this information and thereby enable neural network constitutive modeling in more complex scenarios.

\begin{figure}[h]
	\centering
	\includegraphics[width=0.7\linewidth]{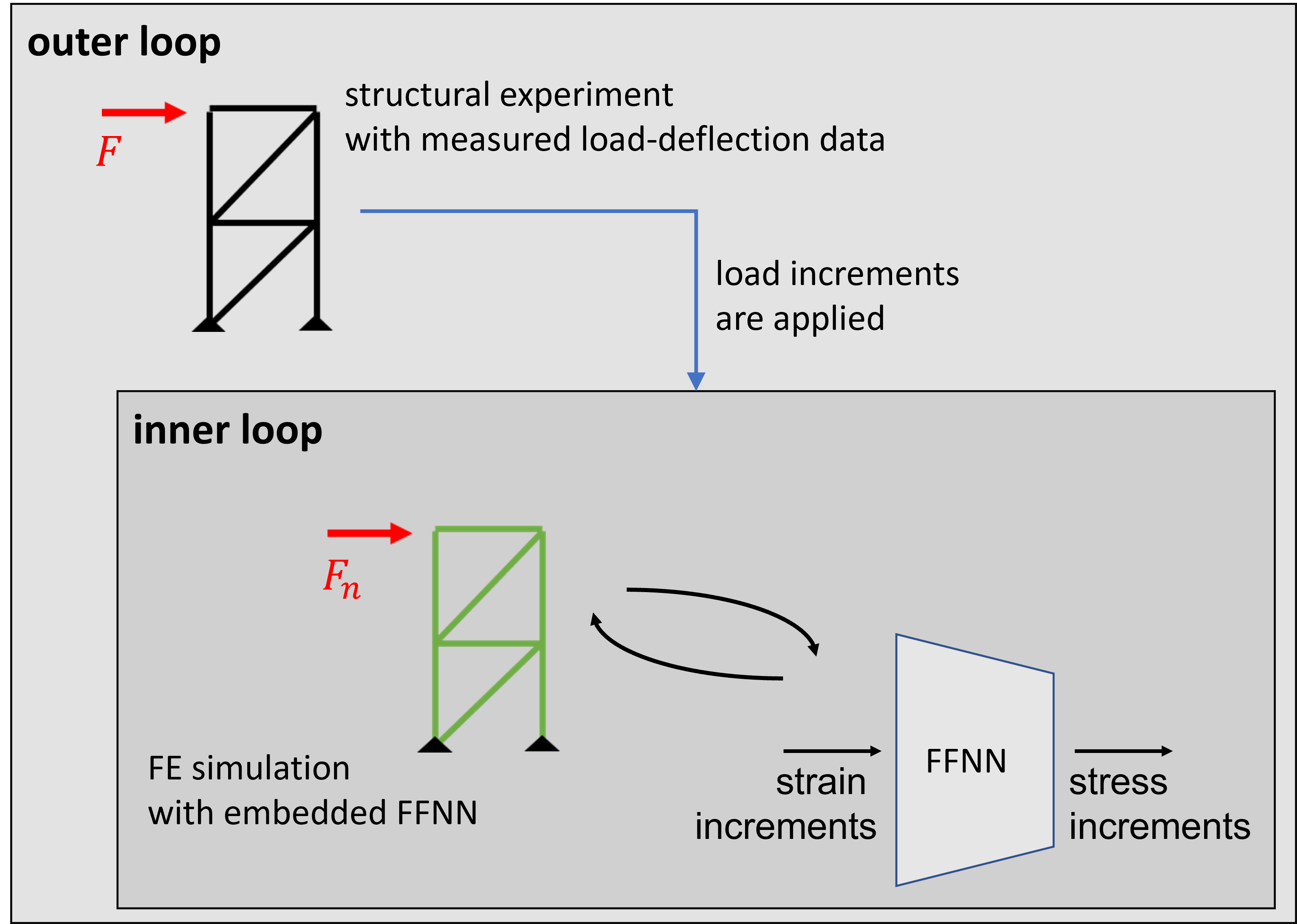}
	\caption{Illustration of training data generation for the autoprogressive training procedure following \cite{ghaboussi1998autoprogressive}}
	\label{fig:autoprogressive_training}
\end{figure}

\begin{figure}[h]
	\centering
	\includegraphics[width=0.7\linewidth]{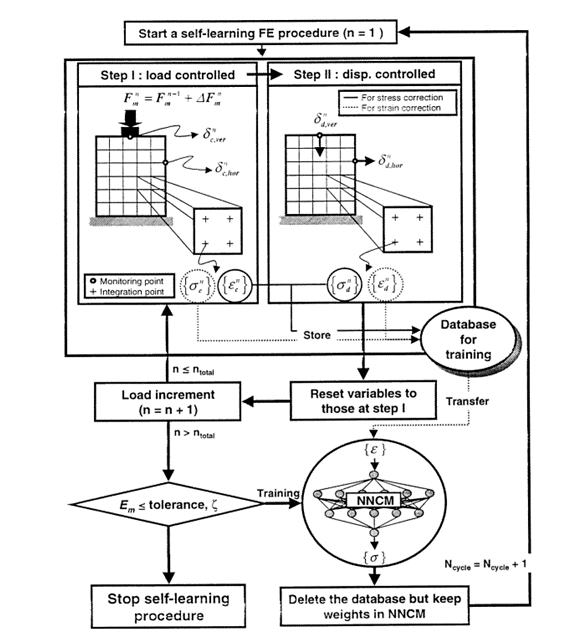}
	\caption{Model architecture of the robust indirect approach introduced in \cite{shin2000self} Copyright\textsuperscript{\tiny\textcopyright} 2000, Elsevier}
	\label{fig:indirect_data_a}
\end{figure}

\begin{figure}[h]
	\centering
	\includegraphics[width=0.7\linewidth]{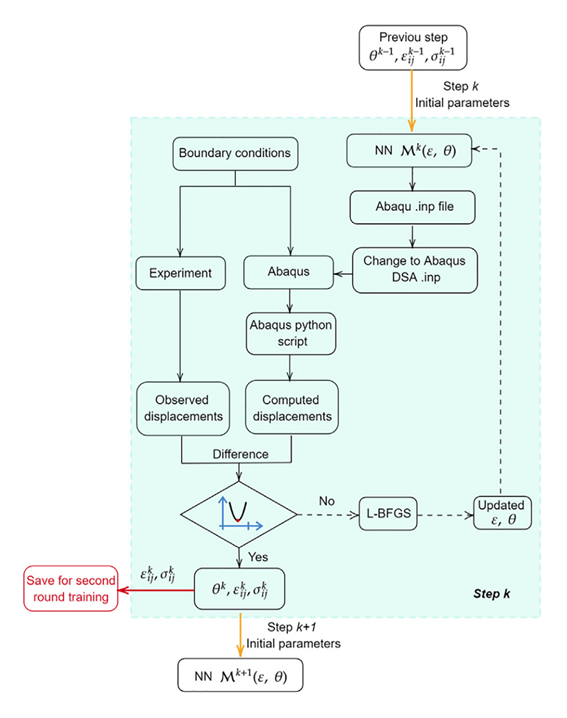}
	\caption{Model architecture of the Abaqus Deep Neural Network approach introduced in \cite{tao2021learning} Copyright\textsuperscript{\tiny\textcopyright} 2021, Elsevier}
	\label{fig:indirect_data_b}
\end{figure}

Besides the direct usage of FFNNs on experimental data for implicit constitutive modeling, Ghaboussi et al. \cite{ghaboussi1998autoprogressive} proposed an early framework for indirect learning. Within the training framework, global load-deflection data from structural tests (i.e. by applying load on a truss structure) are used, combined with an FE simulation of the structural test with an embedded FFNN as constitutive model. After pre-training the FFNN on data from a linear elastic model in the first step, the FFNN is trained in an iterative bootstrapping procedure, which is called autoprogressive training by the authors in \cite{ghaboussi1998autoprogressive}. As depicted in Figure \ref{fig:autoprogressive_training}, each iteration of the autoprogressive training consists of two nested loops to generate training data:
\begin{itemize}
 \item In the outer loop, load increments from the global load-deflection curve are applied to the FE model of the structural component.
 \item For each applied load increment, in an inner loop, the FE model with the embedded FFNN is used to compute the  actual deflection. To meet the measured deflection, displacement corrections (the difference between actual and target deflection increment) are applied in an iterative procedure. As training data for the FFNN, stress-strain data at local material points of the structure are extracted, which follow from this iterative procedure. The inner loop ends, when the stress-strain data converges and the displacement correction is sufficiently small.
\end{itemize}
In \cite{ghaboussi1998autoprogressive}, it is stated by the authors that the procedure described above may need to be performed several times of the full range of the applied load in order to train the neural network material model satisfactorily. The autoprogressive framework is introduced on a simple truss structure example and tested based on a more complex structural setting of graphite epoxy lamina, however with the assumption that the behavior of epoxy lamina is not path dependent. 
Follow-up works that build upon the autoprogressive algorithm apply it to sand under non-uniform triaxial compression \cite{sidarta1998constitutive} and to model soil behavior from field measurements excavations \cite{hashash2003systematic}. Methodical extensions of the autoprogressive algorithm are proposed by Yun et al. \cite{yun2008self}, who focus on hysteretic material behavior and by Shin et al. \cite{shin2000self} to improve the overall robustness of the training framework. 
The autoprogressive framework was expanded by Yun et al. for applications with cyclic loading and hysteretic behavior \cite{yun2008self}. To model such behavior, a neural networks-based material model using a novel algorithmic tangent stiffness formulation was proposed. 

The adopted autoprogressive procedure was applied to a structure of hysteretic beam-column connections \cite{yun2008design} and it was shown that the trained neural network model has a superior learning capability compared to the previous direct neural networks-based material models. Moreover, the neural network was shown to be able to successfully extract the local cyclic behavior from the global responses measured in synthetic and real experiments and is capable to generalize to unseen cyclic motions.
In this context, Shin and Pande \cite{shin2000self} proposed a more robust variant of autoprogressive training. The proposed training scheme is depicted in Figure \ref{fig:indirect_data_a}. A central difference to the original autoprogressive training (\cite{ghaboussi1998autoprogressive}) is that the neural network is retrained only once per load pass by using only data from the current load pass. Also, the strain correction in the original autoprogressive training is complemented with a stress correction scheme to enable the successful training of strain softening. The authors of \cite{shin2000self} investigated furthermore how the location and number of global monitoring points influence data efficiency and prediction accuracy. The autoprogressive framework relies on a proper initialization of the neural network parameters for fast convergence \cite{liu2020learning}, e.g. based on an approximate linear elasticity matrix \cite{shin2000self}. Moreover, as shown in Figure \ref{fig:indirect_data_a}, two FE simulations are needed per iteration. In the so-called stress correction scheme (which is contrary to the strain correction scheme), the first simulation is load controlled and yields a displacement field, which is the basis for the second displacement controlled simulation yielding a stress field. The neural network is trained on the resulting strain-stress pairs until the observed and predicted displacement field match. However, such coupled simulations potentially lead to computational issues for applications with complex structural simulations \cite{tao2021learning}.

In the context of dual phase steels \cite{li2019machine}, in multi-scale modeling \cite{huang2020learning, xu2020inverse, xu2021learning} and laminates \cite{liu2020neural, liu2020learning}, several methods for indirect learning were proposed in recent years. However, as \cite{huang2020learning, xu2021learning} involve the explicit formulation of energy principles, we classify them as physics-integrated surrogate models, which are discussed in Section \ref{physics_indirect}. The other approaches mentioned are discussed in the following. Li et al. \cite{li2019machine} trained a neural network as part of a modified Johnson-Cook model that reflects the non-monotonic strain rate and temperature effects on strain hardening of dual phase steel from experimental tensile test data. As before, the training of a constitutive neural network is accomplished in a bootstrapping process. First, the neural network is initialized by pre-training on user estimations of the material behavior under different temperatures and strain rates. In the second step, the pre-trained network is embedded into a 3D FE simulation and trained in the bootstrapping process, where the cost function is derived from the difference between the predicted and experimentally measured total forces, assuming a linear variation of axial stresses over cross-sectional areas. Although the authors measure the strain distribution on the surface experimentally, this has not been included in the error function for neural network training. The rich local information of strain distribution is lost here, which if used would increase the accuracy and generalization of the FFNN to a heterogeneous field.
Based on the same approach, Pandya et al. \cite{pandya2020strain} trained a model to describe the rate-dependent plastic behavior of aluminum 7075 in a hot stamping process. 

For various examples from lamination, Liu et al. \cite{liu2020neural, liu2020learning} learned constitutive relations based on experimental data by combining neural networks with lamination theory, which is implemented into differentiable FE code. The goal thereby is not to learn a general stress-strain model, but instead more context-dependent models such as shear constitutive relations and the failure initiation criterion \cite{liu2020learning} or the damage accumulation law \cite{liu2020neural} of laminates. Although the results are very convincing, the requirement of differentiability of the used FE code leads to a very elaborated development process in the case of more complex FE problems \cite{tao2021learning}.
In \cite{tao2021learning}, a method is proposed to replace the custom FE code with the commercial Abaqus FEM solver. The training procedure of the so-called Abaqus Deep Neural Network is outlined in Figure \ref{fig:indirect_data_b}. Within this approach, a neural network (with parameters $\boldsymbol{\theta}$)  that maps from strains to the Jacobian of $\boldsymbol{\sigma}(\boldsymbol{\varepsilon})$ 
\begin{equation}
\tilde{f} (\boldsymbol{\varepsilon},\boldsymbol{\theta}) = \frac{\partial \boldsymbol{\sigma}}{\partial\boldsymbol{\varepsilon}}
\end{equation}
is included into the Abaqus FE simulation and is trained based on the difference between displacements observed in experiments and displacements computed by the FE simulation. The central contribution of \cite{tao2021learning} is the utilization of the so-called Abaqus design sensitivity analysis functionality to compute the gradient of the computed displacements with respect to the constitutive neural network and the integration of this functionality into backpropagation. The proposed method was applied to learn the progressive damage constitutive law of a fiber-reinforced composite and the linear constitutive laws of its constituents based on structural-level data for a specific loading scenario.

The indirect learning approach using FFNNs has rarely been employed to tackle path-dependent elasto-plastic or plasticity-driven ductile damage problems. The auto-progressive method, while intriguing, demands significant computational resources and necessitates expertise in both, FEM and constitutive modeling. In their work, Li et al. \cite{li2019machine}, present an framework for adaptively identifying the flow stress but make the assumption that the plastic flow behavior is already known. It would be interesting to explore the entire elasto-plastic flow behavior in this context.

\subsection{Discussion}
\label{sec:critique_universal_func} 

As shown in this section, the use of FFNNs to learn constitutive models from experimental or simulated data is very popular. However, due to long-term effects, the state and behavior (e.g. hysteretic) of materials often depend on the load history. Various methods to represent the current material state for FFNNs were proposed and discussed. Examples, as discussed, include the use of past values from sampled isolated time-steps \cite{ghaboussi1991knowledge, lefik2003artificial, hashash2004numerical}, the use of aggregated values \cite{huang2020machine} and the use of additional internal hardening variables \cite{furukawa1998implicit} or the use of processing time-step \cite{al2006prediction}.

However, neural networks are limited to fixed input mappings, while the load history is of variable length. Methods that face the problem by aggregating the history through subsampling \cite{ghaboussi1991knowledge, lefik2003artificial, hashash2004numerical} or using accumulated values \cite{huang2020machine} typically accept some degree of information loss by making assumptions about the history of internal variables. A first attempt in the constitutive modeling context to solve this problem by using a more flexible neural network architecture is the NANN approach. However, the NANN approach is limited to short sequences, as the number of model parameters highly depends on the sequence length. In contrast, advanced modeling techniques like RNNs and time convolution approaches are promising alternatives to solve these issues as in the case of RNNs the number of model parameters is independent from the sequence length and in the case of time convolution, the number of parameters required for learning temporal patterns is greatly reduced.

Furthermore, among the papers reviewed in this section that focus on direct learning, many address the learning of flow stress curves for different materials. Learning such curves is a 1D regression problem and can actually be solved by comparatively simple neural networks. While applying such is a reasonable proof-of-concept for the applicability of using neural networks in materials modeling, it, however, raises the general question of when applying machine learning is advantageous to classic models. In this specific case, an advantage of neural networks is their characteristic of being able to be fitted to any flow curve. However, a huge advantage in execution time is not to be expected. In contrast, fitting one out of a plethora of already developed analytical functions should result in similarly good curve approximation while at the same time forming derivatives is much more straightforward. In other cases, like in higher-dimensional regression problems, in contrast, the use of machine learning is more beneficial as the effort to find analytical functions increases with an increase in dimensions, see for example the modeling of complex yield functions in \cite{hartmaier2020data} and \cite{shoghi2022optimal}.

\section{Advanced Neural Networks for Constitutive Modeling: Considering the load history by learning from time-series}\label{advanced_recurrent}

In this section, we outline and discuss work that utilizes and introduces advanced neural network models for processing sequential data in the context of constitutive modeling. As we are not aware of published indirect approaches that utilize such advanced neural networks, we discuss approaches for direct learning of advanced neural networks for processing sequential data in the following Section \ref{advanced_recurrent_direct} only. In Section \ref{sec:critiques_advanced_recurrent}, a brief discussion of the presented approaches is given.

\subsection{Direct Learning}\label{advanced_recurrent_direct}

In an early series of works by Oeser and Freitag \cite{oeser2009modeling}, Graf, Freitag et al. \cite{graf2010recurrent, graf2012structural}, and Freitag et al. \cite{freitag2011recurrent}, RNNs are applied to constitutive surrogate modeling tasks.
In \cite{oeser2009modeling}, RNNs are proposed as part of fractional material models to learn the history-dependent stress state of rheological materials with fading memory, where the effects of applied stress states on structural behavior gradually diminish over time. The RNN model is applied to synthetic creep test data, for which the better run time behavior in comparison to the exact solution of a fractional differential equation is highlighted. 
Based on this work, in \cite{graf2010recurrent, graf2012structural, freitag2013material}, RNNs are combined with fuzzy structural analysis and the so-called $\alpha$-level optimization \cite{moller2000fuzzy} for history-dependent structural and constitutive models, which reflect measurement and process uncertainties. 
In \cite{graf2010recurrent}, an RNN is used to map from loads and environmental influences of a reinforced concrete plate to structural responses. Both, inputs and outputs, of the RNN are fuzzy variables to reflect the uncertainty of the measurements, in the loading process itself and in the environmental influences. The approach is evaluated on the long-term responses of a reinforced concrete plate under dynamic loading.

%\subsection{Surrogate Constitutive Modeling}\label{advanced_recurrent_surrogate}
While in \cite{graf2010recurrent} models reflect the material behavior on the structural level, in the follow-up work \cite{graf2012structural, freitag2013material}, the approach is applied to learn constitutive surrogates for fuzzy FE models \cite{rao1995fuzzy}. The approach is applied to learn a surrogate of a linear elastic constitutive model in \cite{freitag2013material} and
from a fractional Newton element to simulate visco-elastic material behavior as well as from a fuzzy FE analysis of a three-dimensional structure under long-term loading in \cite{graf2012structural}. 

\begin{figure}
	\centering
	\includegraphics[width=0.7\linewidth]{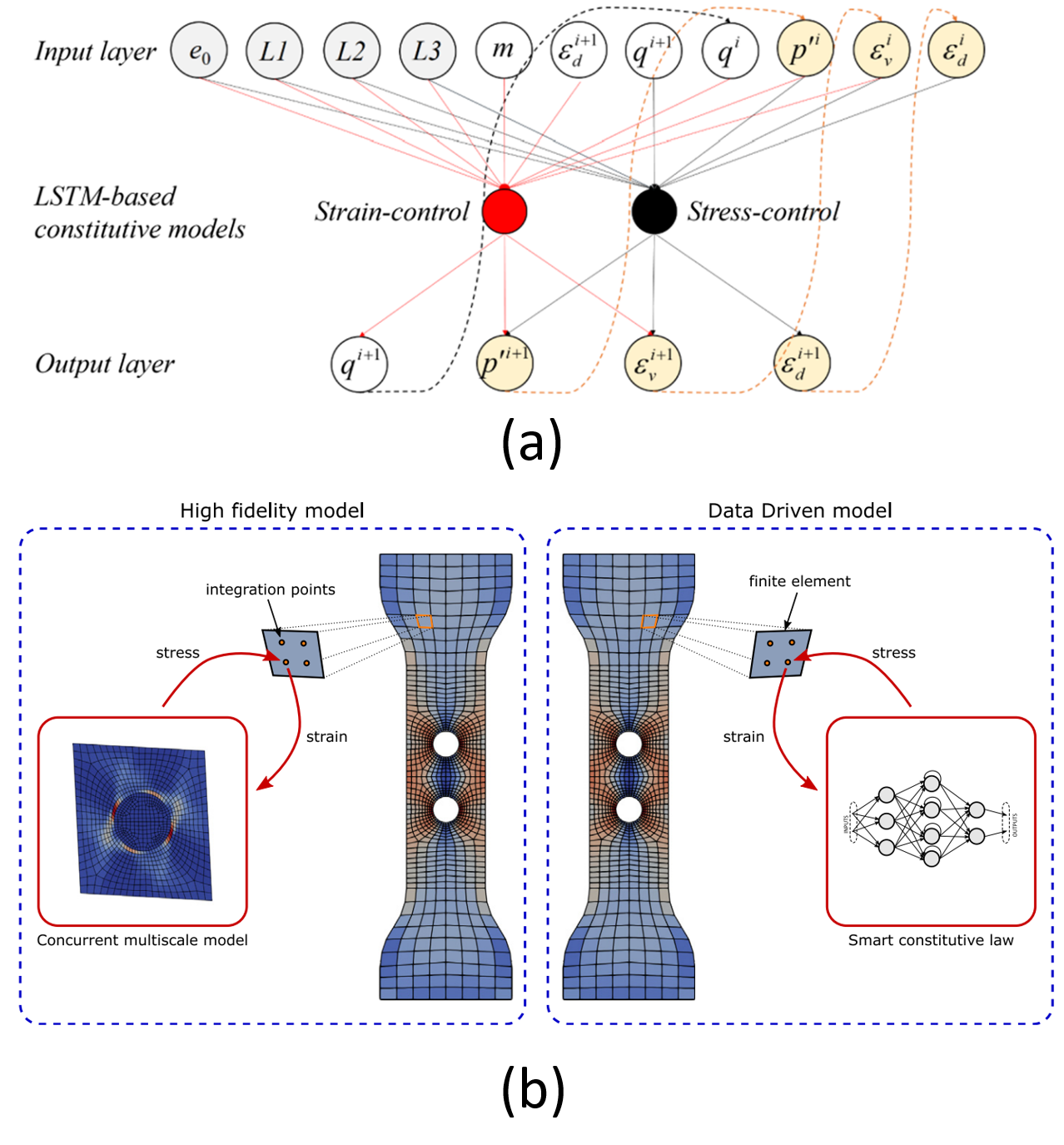}
	\caption{Sequential data processing constitutive neural networks: (a) LSTM-based models for describing cyclic behavior of granular materials \cite{zhang2020ai} Copyright\textsuperscript{\tiny\textcopyright} 2020, John Wiley and Sons, (b) so-called smart constitutive law approach introduced in \cite{logarzo2021smart} Copyright\textsuperscript{\tiny\textcopyright} 2021, Elsevier.
	}
	\label{fig:sequence_models}
\end{figure}

For various single-scale  applications, different types of RNNs have been proposed \cite{gorji2020potential, zhang2020ai, bonatti2021one, abueidda2021deep}. 
Gorji et al. \cite{gorji2020potential} proposed and evaluated GRUs as a substitution for conventional physics-based plasticity models in the large deformation regime. The model is trained based on single-element simulations using a Yld2000-2d yield model with homogeneous anisotropic hardening and evaluated in various settings, including arbitrary multi-axial loading paths. Besides the homogeneous studies, the GRU-based model is trained and evaluated on results from a unit cell analysis of a two-dimensional foam. 
Zhang et al. \cite{zhang2020ai} proposed an LSTM-based approach to model the cyclic behavior of sand under drained and undrained conditions. The proposed model consists of two separate LSTM networks. One network mimics strain-controlled and the other mimics stress-controlled soil behavior. The input and output variables of the proposed models are depicted in Figure \ref{fig:sequence_models} (a), where $p^i, q^i,\epsilon_v^i, \epsilon_d^i$ are the mean and deviatoric stresses, volumetric and axial strains at time step $i$ and additional variables $L1,L2,L3$ describe the current loading stage. Further, the features $m$ and $e_0$ were introduced, to describe the drainage and the initial void ratio of sand. In the presented study, the model is trained on synthetic data of drained and undrained sand and is employed to simulate the behavior of real sands under cyclic loading. The model is evaluated on experimental data. As part of the evaluation, the trained model is shown to be able to accurately predict history-dependent effects such as shear strain accumulation and densification. 

More recently, Zopf and Kaliske \cite{zopf2017numerical} proposed to combine constitutive neural networks and RNNs with the so called micro-sphere description \cite{miehe2004micro} to model the finite strain behavior of rubber-like materials. The micro-sphere model enables the reduction of the stress-strain dependency of polymer chains to only one dimension and thereby allows to train constitutive models with data from uniaxial loading tests to drastically reduce the experimental effort. The proposed model consists of both, an FFNN for pure elastic behavior and an RNN for inelastic behavior. The model is implemented into an FE model which is evaluated on experimental data of uncured elastomers. 

Bonatti and Mohr \cite{bonatti2021one} proposed a special recurrent architecture for constitutive sequence models, which is designed to combine a high model capacity with an arbitrarily small material state representation. The recurrent architecture is used to learn a general stress-strain model and consists of special quadratic layers followed by a simplified LSTM layer, which learns an implicit state representation of the material state. The state variables are connected to the next time step input and gated layer. Various separate constitutive surrogates are trained on the basis of the proposed architecture on data from FE simulations of single elements. Besides isotropic-hardening and mixed-hardening elasto-plastic models, the evaluation studies include models of crushable foam and hyper-elastic rubber with internal damage. Besides the evaluation of the reproduction quality, a study of the correlation between the implicitly learned state variables and the state variables of the conventional models shows the ability of the LSTM-based model to learn physically meaningful state spaces. Recently, Bonatti et al. \cite{bonatti2022importance} published an enhanced version of the model and showed that it can be used as an efficient and accurate constitutive surrogate for FE simulations. 

Abueidda et al. \cite{abueidda2021deep} studied and compared various sequence models to predict the strain and temperature history-dependent behavior in two different applications. LSTM, GRU, and time CNNs are trained on data from an elasto-plastic cellular periodic material and a more complex thermo-visco-plastic steel solidification problem. In contrast to other sequence constitutive models, which are supposed to be implemented into an incremental FE solution procedure, in \cite{abueidda2021deep}, the neural networks are used to predict the entire response sequence in one pass.

Up to this point, we discussed methods that learn path-dependent material models at the macro-scale, where the model represents the material in a homogenized sense. In the following, however, we discuss a second group of approaches namely models that learn the behavior of the material from RVEs, wherein the microstructure is spatially discretized. In computational homogenization, the computational bottleneck is the evaluation of the micro-scale model per integration point \cite{ghavamian2019accelerating}. Due to various microscopic effects, the stress-strain relationship on the macro-scale is typically history-dependent \cite{logarzo2021smart}. RNN-based constitutive surrogate modeling approaches are recently proposed to solve these issues in a data-driven manner \cite{wang2018multiscale, ghavamian2019accelerating, wu2020recurrent, logarzo2021smart}.
Wang and Sun \cite{wang2018multiscale} integrated LSTM networks on different length scales into numerical models and investigated the computational efficiency of this hybrid approach using a multi-scale hydro-mechanical simulation of a multi-permeability porous material. 

A popular approach is to utilize RNNs to learn a surrogate for homogenizing RVEs \cite{mozaffar2019deep} and use the trained model as constitutive law in macro-scale simulations \cite{ghavamian2019accelerating, wu2020recurrent, logarzo2021smart}. 
Mozaffar et al. \cite{mozaffar2019deep} proposed this approach first and used RNNs to learn homogenization models for elasto-plastic composite materials. Therein, the RNN model is trained to map from the deformation path and microstructural descriptors to homogenized stresses and a quantity that the authors called plastic energy. The method is applied to two types of composites, for which it is shown to be accurate and efficient.
Ghavamian and Simone \cite{ghavamian2019accelerating} trained LSTM-based RVE surrogates to accelerate FE$^2$ simulations of history-dependent materials and applied it to virtual strain-softening experiments with a Perzyna-type visco-plasticity description.
In contrast, Wu et al. \cite{wu2020recurrent} proposed GRU-based models as basis for meso-scale surrogates for an FE$^2$ simulation of a fiber reinforced composite under random cyclic loading. 

Further, Logarzo et al. \cite{logarzo2021smart} introduced a so-called smart constitutive law approach, in which an LSTM is trained as a surrogate model on RVEs consisting of an elasto-plastic material matrix with hard elastic inclusions. Besides the homogenized stress components, the LSTM maps to the localized (maximum) plastic strain and von Mises stress. This localized information can be used in the application context, e.g. for predicting premature failure. The trained surrogate model is used as constitutive model in a structural FE analysis, as is depicted in Figure \ref{fig:sequence_models} (b). Using the FE simulation, the machine learning-based approach is compared to a concurrent multi-scale scheme, in which the microstructural boundary value problem is solved explicitly for each integration point. The surrogate model was shown to be very accurate, as the coefficient of determination (R2-score) at the last of the 200 applied load increments was relatively high (above $0.989$) for all stress components. Despite parallelization of the concurrent microstructure level calculations over 36 computing cores, the multi-scale scheme needed over three days to complete, while it took the machine learning-based approach less than 19 minutes to finish on a high-end GPU.

To model path-dependent elasto-plastic crystal behavior, Heider et al. \cite{heider2020so} proposed special RNNs combined with a reference frame invariant formulation of the loss function. Based on synthetic data generated from crystal plasticity simulations, it was shown that the model results are highly dependent on the choice of the input representation and the loss function. It was furthermore shown that accurate results can be achieved with the right choice of the proposed frame-invariant loss in combination with some of the also proposed graph representations of the input data. Albeit the model was proven for a geometrically linear small strain case.

\subsection{Discussion}
\label{sec:critiques_advanced_recurrent}

Generally, for training RNN models, a sufficient amount of various non-linear strain paths and corresponding stress values are necessary. All of the reviewed papers present their own way of generating training data, which is sufficient for proving the introduced approaches. However, it is challenging to determine the superiority of one method over another. Therefore, it is crucial to establish benchmarking problems to evaluate the effectiveness of each method accurately. As the possibilities to sample strain paths is unlimited, appropriate approaches need to be developed in order to provide comparable data sets. These can for example be to limit sampling on sub-spaces for applications in certain domains (e.g. metal forming) or to apply adaptive data generation and model training. Such sampling approaches furthermore enable the generation of standardized data sets that facilitate benchmarking. 
In contrast, another possibility for training models is the application of indirect learning. However, we have not found an indirect learning approach for RNN models in the reviewed literature. Learning indirectly from experiments would be an important extension to existing approaches, also tackling the data generation issue mentioned above.

In addition for path-dependent problems, the time step in FEM plays a crucial role in determining the accuracy and stability of the numerical simulation. Generally, irrespective of the path dependency, a small time step allows for better accuracy, but increases the computational cost of the simulation. On the contrary, a large time step can lead to numerical instability and loss of accuracy. In addition, in process simulations for example, variations in contact conditions necessitate different time step requirements. The path dependency amplifies this issue further, and hence is of importance. Commercially available FE solvers have the capability to automatically adjust the time step based on the system's state to optimize computational costs and maintain model accuracy. Except for \cite{bonatti2021one,bonatti2022importance}, there is a lack of research on how dynamically changing the time step affects RNNs. There are not many robust methods available to train RNNs in a way that mitigates the influence of the time step. This is a crucial aspect in driving the integration of neural networks into FEM and structural analysis beyond the academic community.

Moreover, while RRNs have been widely used for processing sequential data due to their ability to capture temporal dependencies, they suffer from several caveats. One issue is the vanishing or exploding gradient problem, which hinders the effective propagation of information across long sequences, leading to difficulty in learning long-term dependencies. Additionally, RNNs are computationally expensive and challenging to parallelize. Transformer models which utilize the attention mechanism \cite{vaswani2017attention} have emerged as a powerful alternative to address these shortcomings. Attention allows the model to selectively focus on relevant parts of the input sequence, enabling better capture of long-range dependencies without the mentioned gradient problems. Also, transformers process the entire input sequence in parallel, making them highly efficient and scalable. These properties have made transformer-based models already state-of-the-art in the machine learning domains various, e.g. for natural language processing tasks \cite{wolf2020transformers} and subsequently in other time series processing applications \cite{wen2022transformers}. 

\section{Advanced Neural Networks for Constitutive Modeling: Considering spatial information by learning from images}\label{advanced_conv}

Neural networks-based constitutive modeling approaches discussed so far operate typically without spatially integrating neighborhood relationships. However, in some applications it is crucial to consider the local neighborhood (or the sub-scale structure of a material) to accurately predict local quantities. CNNs are typically used for this purpose, hence we primarily discuss work presenting CNN-based models for constitutive modeling taking into account spatial information in the following. As we are not aware of published indirect learning approaches in this context, we review only approaches for direct learning in the following Section \ref{cnn_direct}. In Section \ref{sec:critiques_advanced_conv}, the presented approaches are discussed briefly.

\subsection{Direct Learning}\label{cnn_direct}

Neural networks with convolutional layers have been proposed, e.g. to map RVE images in 2D \cite{mianroodi2021teaching,yang2020prediction} or 3D \cite{frankel2019predicting, henkes2021deep} or from graph representations \cite{vlassis2020geometric} to homogenized stresses \cite{frankel2019predicting, yang2020prediction}, localized stresses \cite{mianroodi2021teaching}, or the elastic energy functional \cite{vlassis2020geometric}. In the following, we discuss these works in chronological order. 

\begin{figure}[h]
	\centering
	\includegraphics[width=1\linewidth]{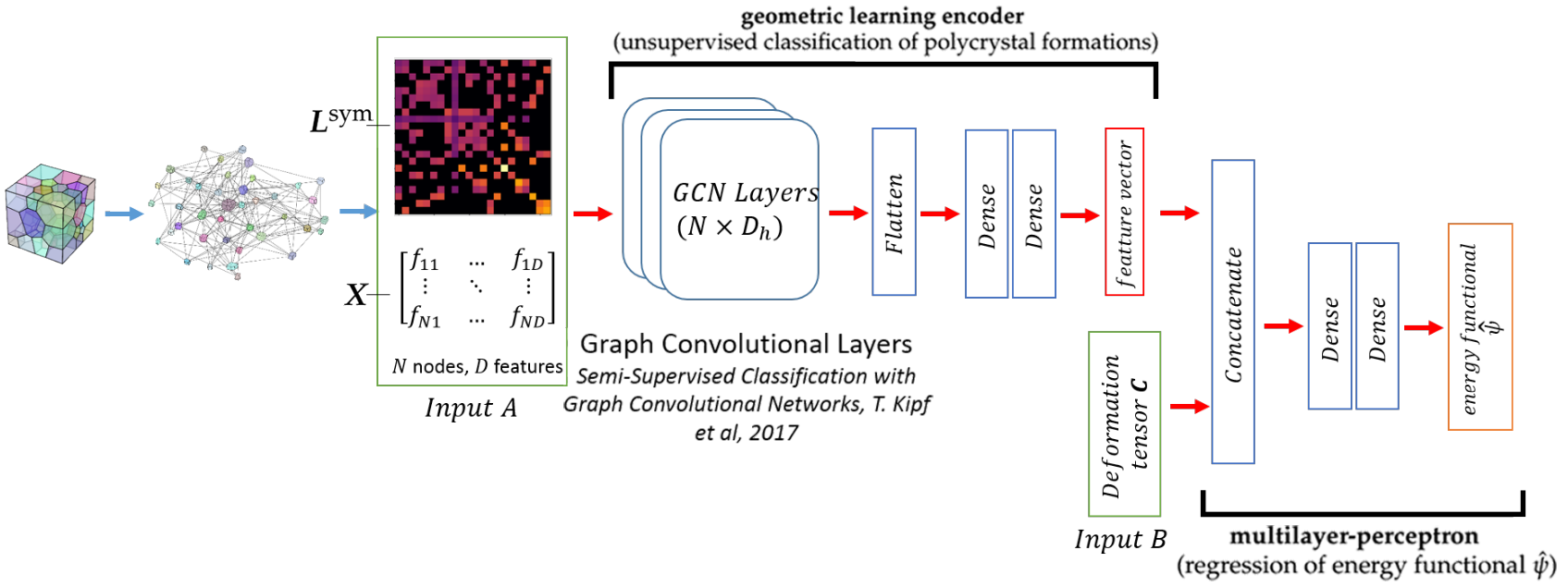}
	\caption{Architecture of the graph convolution model for energy functional prediction of polycrystals introduced in \cite{vlassis2020geometric} Copyright\textsuperscript{\tiny\textcopyright} 2020, Elsevier}
	\label{fig:advanced_conv}
\end{figure}

In his work, Frankel et al. \cite{frankel2019predicting} proposed an RNN with convolutional layers for the first time in order to predict the homogenized stress of oligocrystals from the strain history and a 20x20x20 voxel image of initial grain orientations. 
In another work, Vlassis et al. \cite{vlassis2020geometric} proposed a graph convolution-based neural network to predict the elastic energy functional for polycrystals under deformation. The polycrystal is represented by a graph, in which each node encodes a single crystal with its features, such as e.g. orientation and volume, and the edges encode the crystal connectivity. 
While the representation of crystallographic data as voxelized images is sensitive to the grid resolution and noise, the embedding into graphs followed by graph convolution enables to encode rotational invariances and frame indifference priors into the model. Thereby, convolution-based deep learning can be applied efficiently.

As shown in Figure \ref{fig:advanced_conv}, the geometric encoder consists of graph convolution layers and dense layers and transforms the graph into a feature vector, which is concatenated with the deformation tensor in Voigt notation and processed by dense layers to finally predict the energy functional.
The proposed loss term, furthermore, incorporates the gradients of the reference function with respect to the input based on so-called Sobolev training \cite{czarnecki2017sobolev} in order to not only approximate the constitutive functions output but also its derivatives. This provides additional supervision, which leads to more sample-efficient training. Besides, the accurate stress tensor is obtained, as it is the gradient of the function with respect to the deformation tensor.
In \cite{vlassis2020geometric}, it was shown that the proposed graph-encoding and neural architecture can leverage the crystallographic information in a reference frame indifferent manner and in consistency with the known thermodynamics principles. The usage of Sobolev training to guarantee being consistent with thermodynamics can, however, be seen as incorporating physical knowledge into the neural network learning process, which is a central point of many models that are summarized in the following Section \ref{grey_box}.

Yang et al. \cite{yang2020prediction} proposed a CNN to predict the homogenized stress-strain curve for a composite microstructure under uniaxial loading. The microstructure is represented by a 2D binary image, which encodes the distribution of a soft and a hard phase. The network is trained to predict principal components, from which the stress-strain curve can be obtained by reverting the a priori applied PCA transform. 
Mianroodi et al. \cite{mianroodi2021teaching} proposed a U-net type CNN as a surrogate of spectral simulations of inhomogeneous microstructures. The proposed neural network maps from 2D patches of local material properties (Young's modulus, Poisson ratio, yield strength) to 2D patches of according local von Mises stresses. When trained on 950 samples, the reported MAPE is 3.8\% in the case of elastic and 6.4\% in the case of elasto-plastic materials. Thereby, an immense performance gain compared to the spectral simulations can be observed. In this particular use case, the constitutive neural network approach is 103 times faster than the reference solver for elastic materials and 8300 times faster for elasto-plastic materials.

\subsection{Discussion}
\label{sec:critiques_advanced_conv}

As in Section \ref{advanced_recurrent}, we did not come across any approach that uses indirect learning in the context of CNNs. However, it cannot be concluded that indirect learning is not applicable to the approaches described in this section. For instance, indirect learning may be accomplished in a 2D setting using in-situ EBSD or SEM measurements, followed by a simulation involving a CNN that attempts to match the surface deformation. However, for 3D microstructures, the only viable means of monitoring the 3D deformation field would be through high-energy accelerators, which is seldom accessible. 

Nevertheless, as a first step towards the application of such models to real data, experimental data can be used to evaluate models that are directly learned from synthetic data. However, it is worth noting that all the methods discussed in the reviewed literature pertain exclusively to academic synthetic data. The validation of CNN models using real-world data remains limited. Achieving this is indeed challenging, as synthetic training data needs to be very similar to experimental data (including realistic microstructure representations). However it may be feasible to accomplish this at the surface level using advanced CPFEM in combination with in-situ EBSD measurements.

In general, CNNs are designed to work well on grid-like data, and can therefore be applied to 2D or 3D images of materials. However, there are certain limitations and challenges associated with the usage of CNNs. First, CNNs alone cannot be employed for sequence data processing. This necessitates the integration of other neural network layers, such as RNNs (e.g. for an effective analysis in the case of path dependencies). Another critical aspect of using CNNs for materials data is the computational expense associated with 3D convolutions, particularly with high-resolution 3D data. The data has to be represented as voxelized data, which typically increases the computational burden and memory requirements. To address this challenge, graph-based CNNs have emerged as a promising alternative, particularly when values can be aggregated at the crystal level. By exploiting the local and global connectivity patterns within the crystal, graph-based CNNs can overcome some of the limitations of traditional CNNs. However, it is essential to carefully consider the specific requirements and characteristics of the data at hand when selecting the appropriate model.

\section{Physics Integrated Machine Learning for Constitutive Modeling}\label{grey_box}

This section deals with constitutive modeling based on integrating physics knowledge into the neural network learning process. The advantage of integrating physics is twofold: On the one hand, physics-informed models typically need fewer data to train than pure black-box approaches and, on the other hand, the models can extrapolate in a certain manner (e.g. extrapolation is consistent with integrated physics relations). Direct learning methods are presented in following Section \ref{physics_direct} and indirect learning methods are presented in Section \ref{physics_indirect}. A brief discussion on the presented approaches is given in Section \ref{sec:critiques_grey-box}.

\subsection{Direct Learning}
\label{physics_direct}

From all the literature reviewed, we found only one work that introduces a model with integrated physics relations that was trained on experimental data (however, it can also to be trained on simulation data) \cite{weber2021constrained}. In the approach described therein, the neural network loss function is customized such that predictions do not violate specific physical relations. This is achieved by integrating physical constraints into the learning process by adding penalty terms to the loss function. In the example of learning hyper-elastic material behavior, four constraints are applied particularly, which ensure that (i) the stress components remain zero when no strain is applied, (ii) the conservation of energy holds, (iii) the symmetry of the stiffness matrix for linear elasticity is retained and (iv) the material isotropy for nonlinear elasticity is hold.

% constitutive surrogate
In contrast to \cite{weber2021constrained}, many works exist that train constitutive surrogates. In this regard, we start with the approach described in \cite{hernandez2021deep}, which is similar to the one introduced in \cite{weber2021constrained} but evaluated using simulation data alone. In \cite{hernandez2021deep}, a loss term is added to the neural network loss function to ensure being consistent with thermodynamics by punishing violations of the conservation of energy and the entropy inequality. The presented model was analyzed by training the network on simulation data of an Oldroyd-B-fluid and a hyper-elastic tyre. A different approach is described in Haghighat et al. \cite{haghighat2022constitutive}, which uses PINNs \cite{raissi2019physics} as the basis for a framework to solve material-related differential equations. A  boundary value problem solver incorporating elasto-plastic material behavior and damage formulations was developed and applied to a plate deformation problem. In \cite{haghighat2022constitutive}, the authors stated that their approach can be applied to stress-strain data from meso- or micro-mechanical as well as molecular dynamics simulations. However, the PINN-based framework is only applicable to homogeneous stress and strain distributions.

Other approaches aim to set up models that learn thermodynamically consistent constitutive relations by mapping strain states to corresponding energies and deriving the stress states downwards. Consistency with thermodynamics is commonly obtained by predicting thermodynamic quantities (e.g. energies) instead of predicting stress components, as is done, for example in \cite{xu2021learning}. Earlier works already account for this while applying neural networks to learn hyper-elastic materials behavior, see \cite{shen2004neural,liang2008neural}. In general, hyper-elasticity is suited well for learning, as it is state-dependent and thermodynamics quantities can be approximated using standard FFNNs (without the need for storing internal history variables).

The first method we draw attention to was recently proposed by Masi et al. \cite{masi2020material,masi2021thermodynamics}, who introduced so-called thermodynamics-based neural networks. These are special neural networks that incorporate thermodynamics principles to learn the constitutive behavior of strain-rate dependent processes at the material point. The thermodynamics principles, namely the balance of energy and the dissipation inequality, can be written in the form of the following two equations \cite{masi2021thermodynamics}:
\begin{equation}
	\label{eq:TANN_sigma}
	\boldsymbol{\sigma} = \frac{\partial \psi}{\partial \boldsymbol{\varepsilon}}
\end{equation}
and 
\begin{equation}
	\label{eq:TANN_dissipation}
	D = -\sum_{i=1}^N \frac{\partial \psi}{\partial \zeta_i} \dot{\zeta_i},
\end{equation}
%that need to be held by the neural network, 
with $\psi$ describing the energy potential, $D$ describing the rate of mechanical dissipation and $\zeta_i$ describing kinematic variables.

Basically, thermodynamics-based neural networks consist of two connected FFNNs. The inputs for the first network are values of stress, strain, strain increment, and internal variables at a certain time increment, which are mapped to the increment of the internal variables. These and the strain increment are fed to the second neural network, which predicts the energy potential of the subsequent time increment. On this basis, the dissipation rate, as well as the stress increments (Equations \eqref{eq:TANN_sigma} and \eqref{eq:TANN_dissipation}), are evaluated from the network output using automatic differentiation. The dissipation rate is forced to be positive or zero (depending on the material) during training to be consistent with the first and second laws of thermodynamics. In his work, Masi et al. \cite{masi2021thermodynamics} showed that the approach can be used to model 1D elasto-plastic material behavior with kinematic softening. In the example problem in \cite{masi2021thermodynamics}, it was shown that thermodynamics-based neural networks perform better than pure black-box neural network approaches, such as the ones described in Section \ref{black_box}. Moreover, the results showed that the trained thermodynamics-based neural networks are accurate even for data points that are relatively far from the training range.

Similar to thermodynamics-based neural networks, but with a different neural network structure, Linka et al. introduced so-called constitutive neural networks in \cite{linka2021constitutive}. The inputs for the constitutive neural networks are the right Cauchy-Green tensor and (if necessary) non-kinematic information, such as microstructure or process conditions. On this basis, several sub-neural networks learn to predict invariants of the right Cauchy-Green tensor and the structure tensor. These are used to predict the strain energy in a subsequent neural network. The strain energy is in-turn used to derive the stress and stiffness tensor via automatic differentiation (as is done for the thermodynamics-based neural networks mentioned above). The applicability of constitutive neural networks to engineering problems is furthermore discussed in \cite{linka2021constitutive}. Therein, two use cases are analyzed, which are (i) predicting effective material properties of a matrix-inclusion composite RVE and (ii) accelerating macroscopic FE simulations of hyper-elastic material behavior. It was shown that the implementation of the trained constitutive neural network models into FE software is simple and computationally efficient. Moreover, it was shown that constitutive neural networks are data efficient and the trained models are robust against noise.

Another approach that is based on thermodynamics formulations is the so-called thermodynamics-informed neural network, introduced by Vlassis \& Sun \cite{vlassis2021sobolev}. In contrast to thermodynamics-based neural networks and constitutive neural networks, only a simple FFNN is used therein, however, Sobolev training \cite{czarnecki2017sobolev} is applied to integrate knowledge about derivatives in the loss function of the neural network (similar to \cite{vlassis2020geometric}, which is already mentioned in Section \ref{advanced_conv}). The loss function is customized, such that a mapping from strain values to hyper-elastic energy functionals can be learned while including constraints for stress and stiffness values. Furthermore, in \cite{vlassis2021sobolev}, it was shown that yield functions and plastic flow can be learned from simulation data of polycrystal RVEs using thermodynamics-informed neural networks. The thermodynamics-informed neural network approach has been developed further for learning rate-dependent, pressure-sensitive plastic behavior in \cite{vlassis2022component}. It is pointed out, in contrast to the models presented in Section \ref{advanced_recurrent}, that this method is better interpretable and allows for a modular design of the material models by for example treating elastic and plastic behavior separately in different networks. As an exemplary engineering use-case, an FE simulation is conducted on the macro-scale that incorporates a constitutive surrogate on the micro-scale trained on FFT simulation data of a polycrystal RVE.

Another neural networks-based approach to mention in this section is the approach described by Fernandez et al. \cite{fernandez2021anisotropic}, which maps the right Cauchy-Green tensor on potential energies to predict hyper-elastic material behavior (similar to the approaches described in \cite{shen2004neural,liang2008neural}). In \cite{fernandez2021anisotropic}, the approach was compared against a neural network that maps the right Cauchy-Green tensor on the corresponding stress components in a deformation problem of a meta material. By comparing the approaches, it was concluded that the approach mapping on potential energies has the best generalization capabilities, which is in accordance with the findings in the works presented above.

\begin{figure}
	\centering	\includegraphics[width=0.8\linewidth]{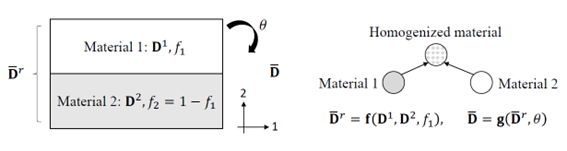}
	\caption{Building block of a DMN and homogenization scheme as is introduced in \cite{liu2019deep}. $\boldsymbol{D}$ denotes the compliance matrix of the individual materials (with volume fraction $f_1$ and $f_2$) and $\overline{\boldsymbol{D}}$ the homogenized compliance matrix for a rotation $\theta$. Copyright\textsuperscript{\tiny\textcopyright} 2019, Elsevier.}
	\label{fig:DeepMatNet}
\end{figure}

A very different approach, compared to the physics-informed approaches mentioned above, are the so-called deep material networks (DMN) \cite{liu2019deep}. DMNs are neural networks that learn homogenization operations to obtain the effective stiffness or compliance tensor from micro-scale simulation data (experimental data is also possible but cumbersome to gather). In contrast to thermodynamically consistent models, DMNs incorporate geometric information from RVEs. In its original version, DMNs consist of building blocks that have a tree-based structure, see Figure \ref{fig:DeepMatNet}. Each building block is a two-layered neural network which gets the micro-scale stiffness or compliance tensor for every phase as input and transforms it to the homogenized quantity, including a rotation operation to incorporate orientation dependence.

In \cite{liu2019deep}, Liu et al. applied DMNs to homogenizing simple two-phase 2D-microstructures. The deep material networks used therein were constructed out of binary tree building blocks. By training the DMNs on data from linear elastic homogenization simulations, it was shown that the approach is able to extrapolate to nonlinear plasticity and finite strain hyper-elasticity. 
For solving 3D problems, Liu and Wu \cite{liu2019exploring} introduced novel building blocks  that include 3D rotation operations and applied them to crystal plasticity among others. Further extensions to DMNs have been introduced in \cite{liu2021cell} to incorporate strain localization enabled by cell division operations and in \cite{liu2020deep} for including interfacial effects via so-called cohesive layers. The ability of DMNs to enable fast multi-scale FE simulations has been shown in \cite{liu2020intelligent} and \cite{gajek2021fe}, both using examples of fiber reinforced composite. 

\subsection{Indirect Learning}
\label{physics_indirect}

In this section, we focus on approaches that learn constitutive relations indirectly with integrated physics knowledge. In particular, all of the models presented in the following integrate physics knowledge aiming to ensure being consistent with thermodynamics.
In this regard, in \cite{xu2021learning}, an RNN-based approach is described that infers the Cholesky factor of the tangent stiffness matrix of a deformed material. The prediction of the tangent stiffness matrix guarantees the approach to be consistent with the second order work criterion, that in-turn guarantees a more numerically stable model.
Training can be done directly using pairs of stress-strain values or indirectly on the basis of measurements by evaluating balance equations during training. However, it is pointed out that measuring comprehensive stress-strain data only from experiments is challenging. The RNN-based model has been shown to successfully approximate and replace elasto-plastic and hyper-elastic material models in 1D and 2D FE simulations. We remark here that the model can be applied to learning in an indirect manner not only, but also in a direct manner.

An alternative approach based on FFNNs that learns constitutive relations indirectly is described in \cite{huang2020learning}. In the approach described therein, neural network models are used to learn constitutive relations by evaluating balance equations (the loss function equals zero when the external forces and the relevant internal quantities are in balance), similar to \cite{xu2021learning}. However, in contrast to \cite{xu2021learning}, the approach described in \cite{huang2020learning} does not require special physics relation to be incorporated in general. In this regard, in the application example, a neural network model is trained to relate principal stretches of a hyper-elastic material model to the first Piola-Kirchhoff stresses. 
It is pointed out that the neural network can be made thermodynamically consistent by predicting the hyper-elastic energy (instead of predicting the stress components) and use automatic differentiation to derive the stress components. However, this alternative has not been applied in the paper.

Besides \cite{xu2021learning} and \cite{huang2020learning}, an early approach, in which FFNNs are trained indirectly using an energy-based loss function has been described already in 2011 \cite{man2011neural}. Therein, neural networks are supposed to learn the relation between stress and strain on a continuum basis. The learning process and the obtained constitutive models are evaluated at FE simulations of structural components that substitute real experiments.

\subsection{Discussion}
\label{sec:critiques_grey-box}

One advantage in learning directly from global experimental stress strain data and indirect learning of constitutive equations is that modeling assumptions and simplifications usually done in classic material modeling do not constrain the learned constitutive model and, hence, make the model more generally applicable. Therefore, constitutive relations need to be learned on the basis of experimental data. Gathering 'enough' experimental data in materials science is, however, demanding, which led to the idea of incorporating physical knowledge in the learning process, aiming to make models more reliable and data efficient. By doing so, the original advantage of universality vanishes depending on how much the machine learning models are constrained by the incorporated physical relations. Consequently, there is a trade-off between improving prediction accuracy and constraining learned constitutive models by incorporating physical knowledge. In this regard, however, we cannot assert our ability to provide recommendations on the appropriate level of physics to be integrated into neural network constitutive models. This depends on the quality and amount of available training data, which is to be assessed by the user.

Additionally, incorporating physical constraints while learning from experimental data is challenging. Although these learning scenarios come from the original intention of learning constitutive models (cf. Figure \ref{fig:application_scenarios}), only few works exist that describe corresponding approaches. The most straightforward approach described in the reviewed papers (and the only one that has been tested on experimental data in a direct learning scenario) is by modifying the neural network loss function in order to punish violations of physical constraints. Indirect learning, in contrast, opens more possibilities for incorporating physical knowledge into the learning process, as the neural network models can learn various relations out of which the material model is composed of (instead of just relating stress and strain). In this regard, few more elaborate approaches have been introduced in the reviewed papers, like inferring the Cholesky factor of the tangent stiffness matrix or using automatic differentiation. It can, however, be expected that there are many more possibilities to decompose the material model and incorporate physical constraints.

\section{Summary and Outlook}\label{conclusions}

The modeling of the constitutive behavior of materials using supervised learning methods, especially neural networks, is a fast-growing research field. As is shown in this review paper, many groundbreaking works have been published in the last years. For example, the modeling of path-dependent material behavior using recurrent neural networks, homogenization using convolutional neural networks, and integrating physics knowledge into the learning process. From a current point of view, the developed approaches are impressive and show the potential of learning algorithms in materials modeling. However, apart from the recent modeling advances, several challenges remain, which in our opinion are:
\begin{itemize}
	\item \textit{Benchmarking}: There is a need for standardized benchmarks to test and compare machine learning approaches for learning constitutive relations against each other. Benchmarks, which are common practice in machine learning (e.g. the MNIST and ImageNet data sets in computer vision \cite{deng2012mnist, deng2009imagenet}), are designed to measure the efficiency of learning approaches and thereby enable comparability. This is necessary for engineers to make the right model decision for their application scenario. Although some of the discussed papers are published with associated code (e.g. in \cite{bonatti2021one, abueidda2021deep, huang2020learning, liu2020neural, fernandez2021anisotropic}) and some related benchmark data sets exist (such as the Mechanical MNIST \cite{lejeune2020mechanical} or MatBench \cite{dunn2020benchmarking}), to the authors knowledge there is no widely used benchmark data set for the comparative evaluation of constitutive neural networks.
	
	\item \textit{Sampling}: The space of possible loading conditions for material models is generally very large. Therefore, as performing numerical simulations and experiments to obtain information about material behavior is often time-consuming, sampling is crucial, see for example \cite{bessa2017framework}. To tackle this issue, intelligent sampling algorithms, e.g. from the field of active learning (cf. \cite{settles2009active}), can be used for data generation and model training, see \cite{rocha2021fly}, \cite{morand2022efficient} and \cite{wessel2022machine} for application examples.
	
	\item \textit{Measuring prediction quality}: Another effect of the large space of possible loading conditions is that learned constitutive models cannot be fitted accurately for any imaginable loading path. Hence, measures to estimate the prediction quality are essential for engineers to assess simulation results. Such measures are given intrinsically when applying probabilistic models, such as Gaussian processes, see for example \cite{bessa2017framework,rocha2021fly}. For descriptive machine learning models, however, such measures have to be still established. One approach can be found in \cite{fritzen2019fly}, where a separate machine learning model is used to estimate the error of a previously trained constitutive surrogate.
	
	\item \textit{Data accessibility}: Having reliable data is the basic requirement for learning adequate constitutive relations. However, data accessibility and management (in particular respecting FAIR-principles \cite{wilkinson2016fair}) is still an open challenge in materials sciences \cite{kimmig2021digital}. Database solutions are already under development (see for example \cite{Himanen2019DataDrivenMS,ekaputra2017ontology,alam2020towards} or relevant websites, such as \url{www.materialscloud.org}, \url{www.nomad-lab.eu} or \url{www.materials-marketplace.eu}) but far away from common use in engineering and industry. In this context, also the fusion of multi-fidelity data sources (e.g. experiments and numerical simulations) is an important research topic, see \cite{batra2019multifidelity}.
	
	\item \textit{Knowledge integration}: Integrating knowledge into the learning process is a prominent way to tackle data sparsity and to improve extrapolation capabilities of machine learning models. Besides the presented ways for knowledge integration, in general, many ongoing research focuses on integrating knowledge into learning processes and thus to create so-called grey-box models, cf. \cite{von2019informed}. Regarding the learning of constitutive equations, additional knowledge can originate from all kinds of balance and conservation relations or physical constraints, such as convexity or monotonicity. First solution approaches that incorporate convexity have been introduced for hyper-elastic material behavior in \cite{klein2022polyconvex}. While for monotonicity no such approaches are known by the authors in the materials modeling domain, however, in applied machine learning, approaches have already been proposed, such as in \cite{kurnatowski2021compensating}. 
	
	\item \textit{Extrapolation}: One of the main limiting factors in machine learning in general is the inability of trained models to extrapolate. The problem of extrapolation can, however, be addressed by methods mentioned in the above presented bullet points. These are: an efficient sampling to prevent from the need to extrapolate, measures for the prediction quality of models to detect and prevent from extrapolation, and, finally, the integration of physical knowledge into the learning process to enable model extrapolation within certain limits.
	
	\item \textit{Interpretability}: A further general issue for using machine learning methods (especially neural networks) is interpretability \cite{du2019techniques}. The lack of interpretability makes it hard for engineers and scientists to analyze and understand model behavior. However, typically, integrating knowledge into machine learning algorithms improves interpretability. Besides knowledge integration, methods are already under development that improve the interpretability and explainability of neural network constitutive model predictions \cite{henkes2021deep, koeppe2021explainable}.
	
	\item \textit{Countering error accumulation}: In numerical simulations, the system response from an applied load is calculated incrementally (stress increments follow from strain increments, etc.). Therefore, errors in stress predictions lead to errors in subsequent increments and so on. 
	The effect of error accumulation is studied in several constitutive neural network papers, including \cite{rocha2020micromechanics, logarzo2021smart,frankel2019predicting}. In \cite{rocha2020micromechanics}, it was shown that error accumulation can be mitigated by regularizing constitutive neural networks. A further way to reduce error accumulation is to incorporate physical knowledge into the machine learning model to force predictions to be consistent with physics constraints.
\end{itemize}
Finally, we want to conclude this review paper by emphasizing that the reviewed works are only the beginning of the era of learned constitutive models. 
While already plenty of approaches exist for building surrogates of classic constitutive models, recently developed approaches that learn from experimental data are rare. Especially for applying deep learning and integrating physics, there is still room for further developments. However, currently, machine learning (and artificial intelligence) research receives a great deal of attention and machine learning methods quickly become replaced by new state-of-the-art methods. At the same time, the model complexity and demand of computational resources is growing. Consequently, this leads to an increased demand for interdisciplinary teams when adapting such models in specific scientific domains such as materials sciences. Further, in materials sciences, as we have seen throughout this paper, model evaluation is particularly difficult, as conducting experiments is often time-consuming and expensive. 

Due to these facts, the adaption of new machine learning methods to problems in the materials science domain takes time. Recent examples are transformer architectures, which are based on the attention mechanism introduced in 2017 \cite{vaswani2017attention} and can consider long-range dependencies in sequential data without the need for recurrence. In constitutive modeling, first results were very recently published showing advantages of transformer models to recurrent neural networks \cite{li2023robust}. Therefore, and in order to push forward the state-of-the-art neural networks-based constitutive modeling and to accelerate development, standard ways for machine learning adaption and implementation into commercial software codes need to be established aiming to facilitate interdisciplinary work. Nevertheless, the speed with which the field of learning constitutive models has evolved in the last decade makes us look forward to new developments in this regard as well as to developments tackling the above mentioned challenges in the near future.

\section*{Acknowledgements}
The authors would like to thank the German Research Foundation (DFG) for funding this work, which was carried out within the research project number 415804944 "Taylored Material Properties via Microstructure Optimization: Machine Learning for Modelling and Inversion of Structure-Property-Relationships and the Application to Sheet Metals".

\section*{Competing interests}
The authors have no competing interests to declare that are relevant to the content of this review article.

\bibliographystyle{ieeetr}
\bibliography{paper}

\end{document}